\documentclass[
reprint,
 amsmath,amssymb,
 aps,
 pra,
]{revtex4-2}

\usepackage{float}
\usepackage{mathtools}
\usepackage{graphicx}
\usepackage{dcolumn}
\usepackage{bm}
\usepackage{upgreek}
\usepackage{lipsum, babel, xcolor}
\usepackage{amsmath, amssymb}
\usepackage{comment}
\usepackage{natbib}
\usepackage{physics}
\usepackage{hyperref}
\hypersetup{colorlinks=true,citecolor=blue}
\usepackage[resetlabels,labeled]{multibib}
\usepackage{standalone}

\newcites{S}{Supplementary References}
\begin{document}

\preprint{APS/123-QED}

\title{Periodically poled thin-film lithium niobate ring Mach Zehnder coupling interferometer as an efficient quantum source of light}

\author{$\mathrm{Mrinmoy\,Kundu^{1}, Bejoy\,Sikder^{1}, Heqing\,Huang^{2}, Mark\,Earnshaw^{2}, A.\,Sayem^{2}}$}

\affiliation{%
 Nokia Bell Labs, USA\\
}%

\date{\today}

\begin{abstract}
Single photons and squeezed light are the two primary workhorses for quantum computation and quantum communication. Generating high-efficiency single photons with high purity and heralding efficiency is the prerequisite for photonic quantum computers. At the same time, generating high-efficiency scalable squeezed light is the prerequisite for continuous variable quantum computing along with sensing applications. Here, we propose a symmetric ring-Mach-Zehnder interferometer (RMZI), which includes a periodically poled lithium niobate (PPLN) waveguide as an efficient source of squeezed light and a single-photon source. We numerically show that our proposed design can generate tunable squeezed light with a squeezing level higher than $\mathrm{-12\,dB}$ with sub-milli-watt (mW) pump power. The proposed device can also generate single photons with purity as high as $\mathrm{99\,(95)\,\%}$ with heralding efficiency $\mathrm{94\,(99)\,\%}$ using only $\mathrm{20\,ps}$ long pulses. Our proposed design is fully compatible with current fabrication technology.

\end{abstract}

\maketitle

\section{Introduction}
Integrated photonic technologies are now at the heart of quantum computation and quantum communication \cite{alexander2024manufacturable,bartolucci2023fusion}, especially with the recent development of scalable on-chip single photon sources \cite{alexander2024manufacturable,qiang2018large,wang2020integrated,ma2020ultrabright,xin2022spectrally,zhao2020high,zhu2021integrated}, switching networks \cite{bartolucci2021switch,bartolucci2023fusion,lenzini2017active,wang2019boson}, and single photon detectors \cite{sayem2020lithium,lomonte2021single,gourgues2019superconducting,cheng2019superconducting,esmaeil2021superconducting,esmaeil2017single,reddy2020superconducting}. High purity single photon sources and single-mode squeezed light are the primary source for photonic quantum computation for linear optical quantum computing \cite{knill2001scheme,kok2007linear,raussendorf2003measurement,briegel2009measurement} and continuous variable quantum computing \cite{menicucci2006universal,gu2009quantum,larsen2021fault} respectively. 
Scalable squeezed light sources with a low pump threshold and a high value of squeezing are required for error-corrected continuous variable quantum computing (CVQC) \cite{menicucci2006universal} and for sensing applications  \cite{lawrie2019quantum}. While squeezed light is the building block for CVQC, single photon sources are the building block for linear optical quantum computing (LOQC) \cite{knill2001scheme}, measurement-based (MBQC) and fusion-based quantum computing (FBQC) \cite{raussendorf2003measurement,bartolucci2021switch}. 
\begin{figure*} [!ht]
    \centering
    \includegraphics[width=1\textwidth]{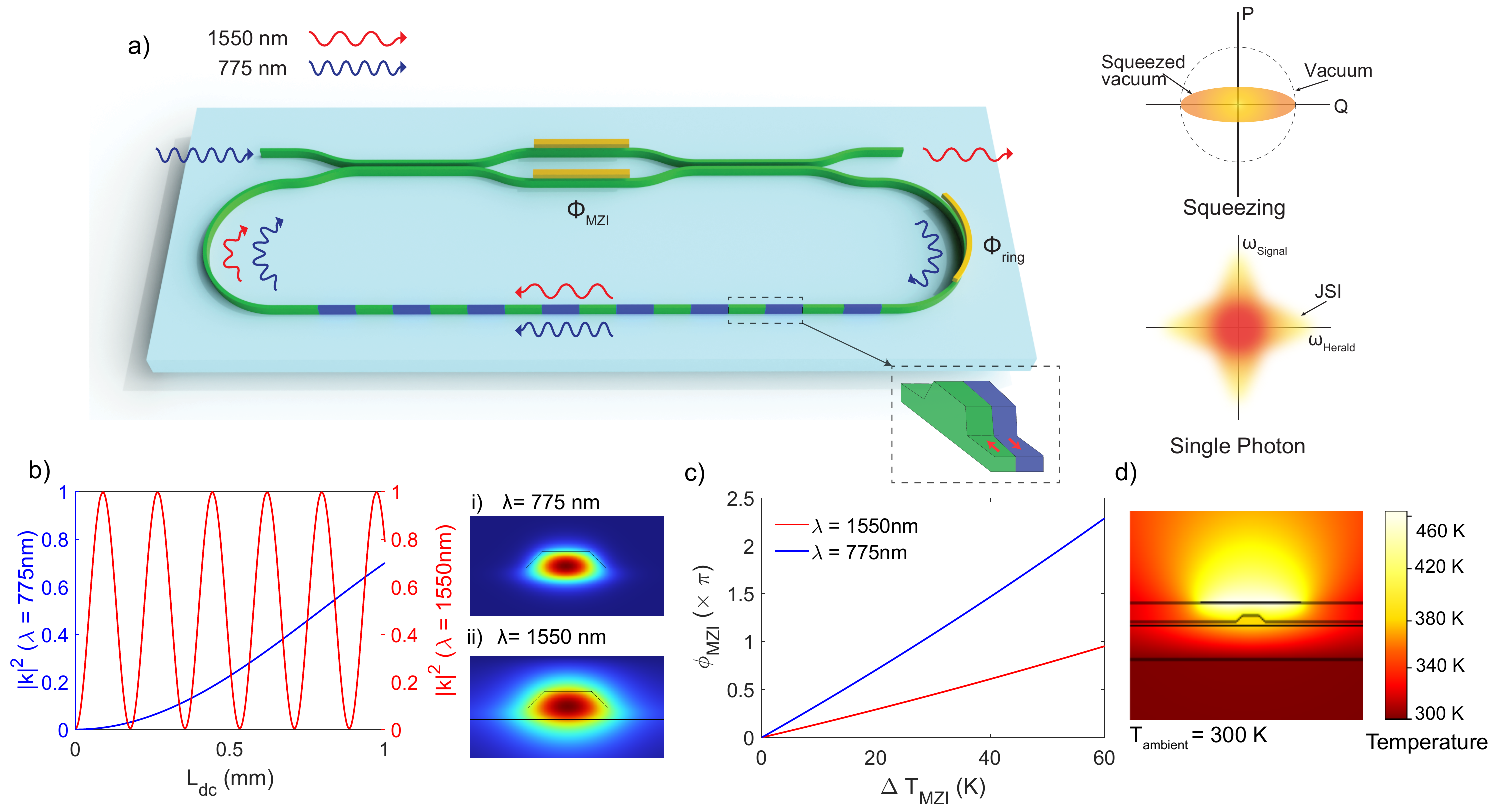}
    \caption{a) Schematic of the PPLN ring resonator coupled with Mach-Zehnder interferometer. b) Simulated cross-transmission coefficient of the directional coupler, $|k|^2$ for g\textsubscript{dc} = $\mathrm{600\,nm}$ at near-IR
    (775 nm) and telecom (1550 nm) wavelengths with varying coupler length $L_{\mathrm{dc}}$. i) and ii) of b) show the fundamental TE mode profiles of the Lithium Niobate (LN) waveguide at these two wavelengths. c) Calculated phase shift, $\phi_\mathrm{MZI}$ as a function of temperature change of the MZI, $\Delta T_\mathrm{MZI}$. d) The temperature distribution of the cross-section of the LN waveguide when heater voltage is applied.}
    \label{fig:Fig1}
\end{figure*}
For the generation of both squeezed light and single photons, either four-wave mixing (FWM) using $\chi^{(3)}$ materials such as Si \cite{paesani2020near,Faruque:18,Wang:18}, SiN \cite{lu2019chip,liu2021high}, AlGaAs \cite{steiner2021ultrabright} or three-wave mixing using $\chi^{(2)}$ materials such as lithium niobate (LN) \cite{zhao2020high,xin2022spectrally,ma2020ultrabright}, AlN \cite{guo2017parametric,liu2023aluminum} etc. can be used. Squeezed light can be generated efficiently through resonant schemes using optical parametric oscillators (OPOs) using either $\chi^{(3)}$ or $\chi^{(2)}$ material platforms \cite{kashiwazaki2020continuous,chen2022ultra,nehra2022few,Meinecke:18}. Due to the ease of spectral filtering and the prospect of cryogenic operation due to the availability of the Pockels non-linearity, $\chi^{(2)}$ materials are preferable compared to the $\chi^{(3)}$ platforms. Especially, $\chi^{(3)}$ non-linearity does not offer low loss and fast electro-optic (EO) active switch networks, hindering the monolithic approach for photonic quantum computation \cite{alexander2024manufacturable}. Among the $\chi^{(2)}$ materials, thin-film lithium niobate (TFLN) is currently one of the most promising and has already been employed in numerous applications such as frequency comb generation \cite{zhang2019broadband,wang2019monolithic}, efficient light generation ranging from mid-IR to UV \cite{sayem2021efficient}, high purity and efficient single-photon generation \cite{zhao2020high}, electro-optic (EO) modulation \cite{wang2018integrated,xu2022dual}, optical parametric oscillator (OPO) \cite{lu2021ultralow}, squeezing light generation \cite{nehra2022few}, wavelength conversion \cite{wang2018ultrahigh} and beyond. LN offers strong $\chi^{(2)}$ non-linearity, fast EO, stable thermo-optic (TO) effect \cite{xu2020high}, and low-loss linear optical characteristics for a broad range of wavelengths necessary for integrated quantum photonic circuits \cite{saravi2021lithium}. 
For MBQC or FBQC, high rate, high purity, and high heralding efficiency single photons are fundamentally crucial for error-corrected quantum computation. Unfortunately, simultaneously achieving all these characteristics is challenging \cite{paesani2020near,xin2022spectrally}. On the other hand, for CVQC, squeezed light is the primary source \cite{konno2023propagating}. Squeezed light has been demonstrated on the TFLN platform but with a low squeezing level\cite{chen2022ultra,park2024single}. On-chip quantum-limited parametric amplifiers can be used to directly measure squeezed light after generation \cite{nehra2022few}. However, such devices require dispersion-engineered waveguides and ultra-fast optical pulses outside the telecommunication band and are incompatible for quantum detection with current photon number resolving detectors \cite{cheng2023100,resta2023gigahertz}.

In this article, we propose a TFLN resonator with a periodically poled lithium niobate (PPLN) waveguide coupled with a symmetric Mach Zehnder interferometer (MZI) to efficiently generate single-mode squeezed light and high-purity single photons. We demonstrate that by using  wavelength-dependent temperature tuning of the MZI and the ring resonator, we can control the coupling condition for the fundamental (FH) and second-harmonic (SH) modes and fulfill doubly resonant conditions to support efficient non-linear optical processes. The variable broadband coupling and resonant conditions can be achieved by thermally tuning the MZI transmission characteristics and the ring resonator, respectively. In the proposed scheme, these two conditions can be almost independently tuned. Thermo-optic tuning has been employed instead of electro-optic because of the compact size of thermal heaters and stable operation \cite{ding2022thermo,wang2022thin,luo2018highly,maeder2022high}. Using our proposed device, we first show that on-chip squeezed light up to $\mathrm{\sim -12\,dB}$ can be generated with pump power as low as $\mathrm{\sim 0.61\,mW}$ by simultaneously fulfilling the over-coupling and the critical coupling condition for the probe and the pump mode respectively. Secondly, we show that using the same coupling condition but with a dual-pulse pump configuration, we can generate single photon pairs with heralding efficiency as high as $\mathrm{95\,\%}$ and with a purity of $\mathrm{99\,\%}$ simultaneously. Our proposed device offers a generalized solution for high-efficiency squeezed light and high-purity single photon generation, which can act as a fundamental building block for both MBQC and CVQC \cite{raussendorf2003measurement,briegel2009measurement}.

\begin{figure*} [!ht]
    \centering
    \includegraphics[width=0.9\textwidth]{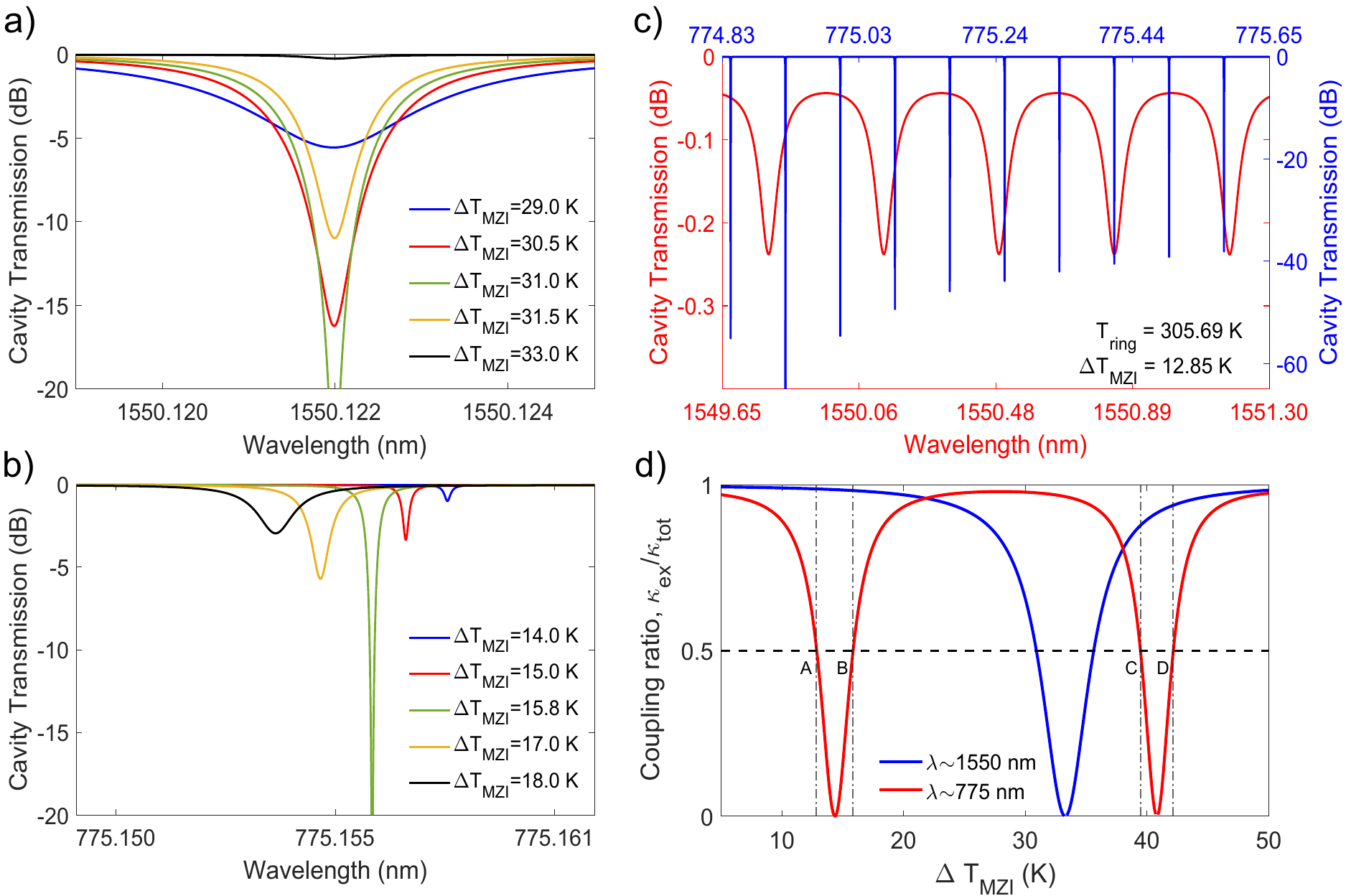}
    \caption{a) The total cavity transmission (in dB) and hence the coupling ratio of the telecom mode can be tuned without varying the resonance wavelength by controlling the temperature change of the MZI, $\Delta T_\mathrm{MZI}$ through thermo-optic effect using the thermal heater. b) Conversely, for pump mode, both the resonance wavelength and transmission are changed with varying $\Delta T_\mathrm{MZI}$. However, the resonance wavelength of the pump mode can be further shifted without changing the transmission by properly setting the ring temperature, $T_\mathrm{ring}$. c) Employing this mechanism, the wavelength detuning of the doubly resonance condition can be tuned to zero. d) The coupling ratio, $\kappa_\mathrm{ex}/\kappa_\mathrm{tot}$ with respect to $\Delta T_\mathrm{MZI}$ for the push-pull configuration. To note, all possible combinations of coupling ratio at near-IR and telecom modes cannot be achieved; however, the required conditions for squeezing light and single photon generation (pump mode is critically coupled and telecom mode is over coupled) are indicated in d) with annotation A,B,C and D.}
    \label{fig:Fig2}
\end{figure*}

\section{Device Concept and Simulation}
Fig.\,\ref{fig:Fig1}(a) shows the schematic of the proposed device. The primary elements of the device are an MZI modulator coupled with a ring resonator connecting the two ports of the MZI. Thermal heaters are positioned at the top of the two arms of MZI and another within the loop waveguide section of the device. The straight part of the loop waveguide section includes a periodically poled lithium niobate (PPLN) to fulfill the quasi-phase-matching (QPM) conditions between the fundamental harmonic (FH) ($\mathrm{\lambda \sim 1550\,nm}$) and second harmonic (SH) ($\mathrm{\lambda \sim 775\,nm}$) modes respectively. Details of the cross-sections of the device with thermal heaters and the directional coupler (DC) can be found in the supplementary section S1\,\cite{supp}. Fig.\,\ref{fig:Fig1}(b) depicts the cross-coupling transmission in the DC for the FH and SH modes. Though a 50:50 splitting ratio is achievable for C-band wavelengths with a coupler length of a few hundred micrometers,  near-IR modes can only render up to $3\%$ cross-transmission at the beat length of the fundamental mode due to the tighter modal confinement. So we choose a larger coupler length for the DC without arbitrarily shrinking the gap between the waveguides. Next, the MZI is configured to work in a push-pull arrangement inducing opposite phase changes in the two arms. This can be achieved by placing either two heaters at the two arms operating at $\Delta T$ and $-\Delta T$ from a base temperature $T_{base}$ or utilizing the electro-optic (EO) effect typically used for high-speed EO modulators \cite{xu2022dual,wang2018integrated}. The induced phase differences between the two arms for our proposed thermo-optic heaters are plotted in Fig.\,\ref{fig:Fig1}(c). Phase differences up to $2\,\pi$ can be achieved with a temperature difference of $\sim 50\,\mathrm{K}$ between the MZI arms. 
Both the MZI and the ring resonator have been considered to be constructed of TFLN waveguides with identical dimensions, e.g., identical width. A detailed derivation of cavity transmission spectra of the proposed device can be found in the supplementary section S2 \cite{supp}. To match the momentum between the FH and SH modes, a PPLN region is considered in the straight section of the resonator. By using the thermal heater on the ring,  the doubly resonance condition can be readily achieved, as shown in Fig.\,\ref{fig:Fig2}(c). Satisfying the momentum and energy conservation requirements, we considered the following Hamiltonian for the degenerate three-wave mixing process in the resonator,
\begin{equation}
\label{Eq_H}
\mathcal{H}=\omega_{a} \hat{a}^{\dagger} \hat{a}+\omega_{b} \hat{b}^{\dagger} \hat{b}+g_0\left(\hat{a}^2  \hat{b}^{\dagger} +(\hat{a}^{\dagger})^2 \hat{b}\right)
\end{equation}
where, $\mathrm{\hat{a}} (\mathrm{\hat{a^{\dagger}}})$ and $\mathrm{\hat{b}} (\mathrm{\hat{b^{\dagger}}})$ are the annihilation (creation) operators for the two modes at a frequency $\mathrm{\omega_{a}}$ and $\mathrm{\omega_{b}}$ respectively where $\omega_b = 2\omega_a$, and $g_0$ is the vacuum coupling rate which depends on the second-order nonlinear coefficient $\chi^{(2)}$ of LN as well as the modal overlap and the geometry of the device. We calculate the $g_0/2\pi$ to be $0.198\,\mathrm{MHz}$ with the PPLN region covering 45\% of the total length of the RMZI resonator. We note that the Hamiltonian in Eq.\,\ref{Eq_H} only describes the $\chi^{(2)}$ interaction between the modes inside the resonator. However, the complete dynamics of the full RMZI PPLN device include intrinsic loss mechanisms and extrinsic couplings through the input and output ports of the upper arm of the MZI. The delicate balance between the coupling ratios defined by the ratio of the intrinsic and extrinsic loss rate of the FH and SH modes determines the performance of the proposed device for the generation of squeezed light and as well as single photons. Here, we denote the coupling ratio as $\eta_i = \frac{\kappa_{i,\mathrm{ex}}}{\kappa_{i,\mathrm{tot}}}$, where, $\kappa_{i,\mathrm{ex}}$ and $\kappa_{i,\mathrm{tot}}$ are respectively extrinsic and total loss rate for mode $i = \mathrm{\{a, b\}}$.

The MZI modulator has been used to tune the transmission spectra of the ring resonator to achieve suitable coupling conditions for both the telecom and near-IR photons. As shown in Fig.\,\ref{fig:Fig2}(a), this configuration enables amplitude tuning of the MZI transmission spectra without varying its phase, ensuring that only the magnitude of the transmission spectra of the ring resonator changes while maintaining the resonance wavelength constant throughout the tuning process. However, due to asymmetry in the cross-transmission coefficient for the near-IR mode, the imbalanced couplers lead to a slight frequency shift as shown in Fig.\,\ref{fig:Fig2}(b). Though a wide range of coupling conditions for both modes can be achieved with proper tuning of the MZI heater, these coupling conditions cannot be tuned independently. Since the thermo-optic heater is shared between the two modes, it introduces correlated phase differences. Hence, it hinders the possibility of independently controlling the coupling conditions with only a single tuning knob, $\Delta T_\mathrm{MZI}$. However, as shown in Fig.\,\ref{fig:Fig2}(d), a finite mismatch between the thermo-optic coefficients of TPLN at near-IR and telecom modes allows us to find suitable coupling conditions for modes around the resonant wavelengths within $\sim 45\,\mathrm{K}$ of temperature tuning for the push-pull configuration. It should be mentioned that coupling ratios remain effectively constant for all other resonant modes both for telecom and near-IR regime at any particular MZI tuning point. This is attributed to the fact that MZI has symmetric arms, and waveguide dispersions for the modes are relatively small. In Fig.\,\ref{fig:Fig2}(d), four characteristic MZI tuning points are also labeled, which shows the capability of the device to obtain highly overcoupled conditions for the telecom mode and critically coupled conditions for the near-IR mode. It is important to note these four characteristic MZI tuning points are the consequences of the particular MZI configuration and it's in general always possible to find these tuning regimes within different fabrication tolerances (e.g. dimension and loss). 
Fig.\,\ref{fig:Fig2}(c) and Fig.\,\ref{fig:Fig2}(d) show that we can satisfy both the doubly resonance and desired coupling conditions using the two temperature tuning knobs namely, $\Delta T_\mathrm{ring}$ and $\Delta T_\mathrm{MZI}$.\\

\section{Efficient Squeezed light generation}
\begin{figure*}[t!]
        \centering
    \includegraphics[width = 0.8\textwidth]{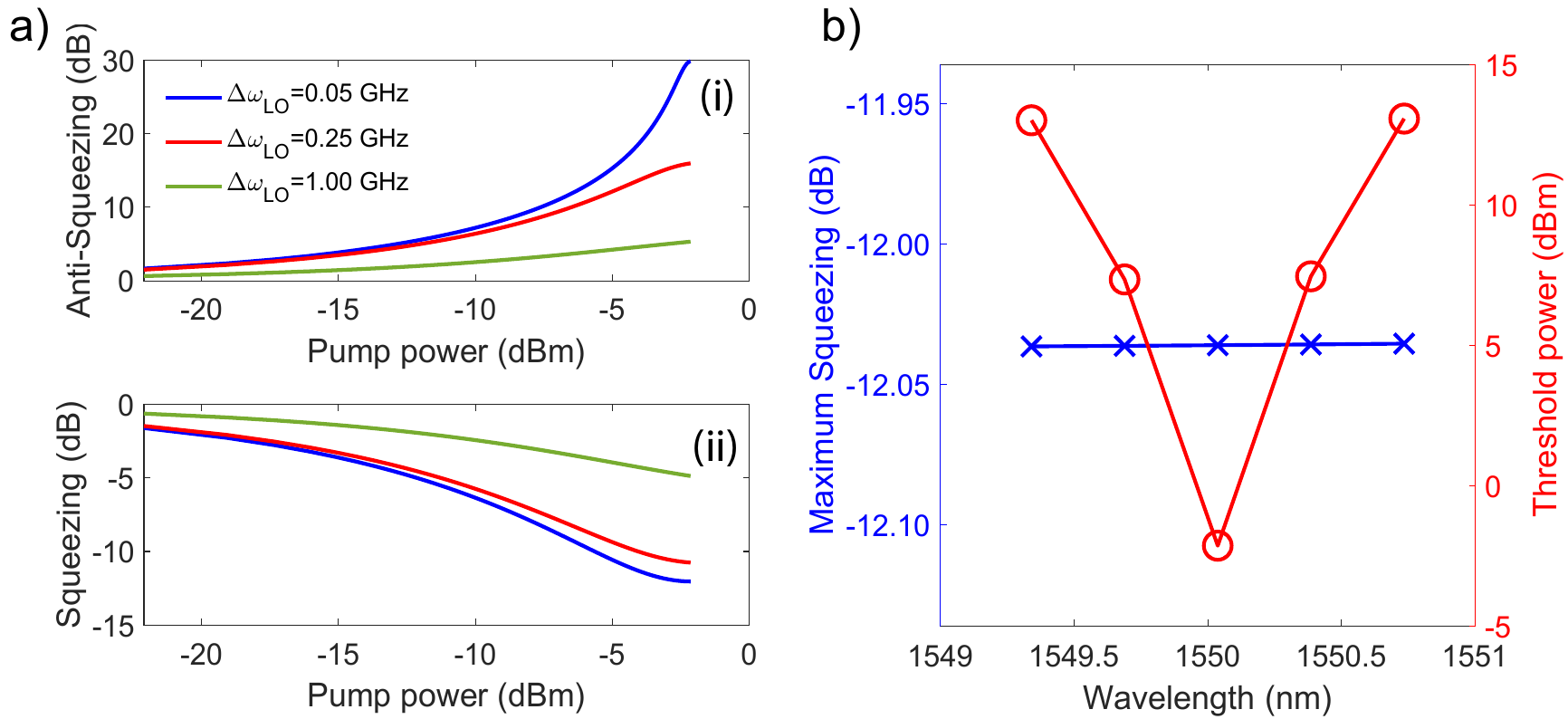}
    \caption{a) (i) Anti-squeezing and (ii) squeezing level as a function of input pump power, $P_\mathrm{s}$ for different de-tunings $\Delta\omega_\mathrm{LO}$ of the local oscillator. At a de-tuning of $50\,\mathrm{MHz}$, a squeezing level of $-12$\,dB can be achieved with an input power of  $0.61$\,mW. Both anti-squeezing and squeezing levels decrease with an increase in the oscillator de-tuning. (b) Maximum squeezing (dB) and threshold power, $P_\mathrm{th}$\,(dBm) for different resonant wavelengths when the energy matching condition is only satisfied at 1550\,nm. It is noted that as the de-tuning from the resonant wavelength increases, the maximum attainable squeezing level (dB) remains invariant while the required $P_\mathrm{th}$ increases quadratically (up to $10\times$) when the wavelength is slightly varied within $\pm 1$\,nm.}
    \label{fig:Fig4}
\end{figure*}

For the efficient generation of squeezed light, we employ the proposed RMZI-PPLN device as depicted in Fig.\,\ref{fig:Fig1}(a) to act as a doubly resonant optical parametric oscillator (OPO). Here, the output mode at $\mathrm{\lambda \sim 1550\,nm}$ is termed as mode $a$, and the pump mode at $\mathrm{\lambda \sim 775\,nm}$ is termed as mode $b$, indicated by the red and blue color in Fig.\,\ref{fig:Fig1}(a) respectively. The OPO consists of the interferometrically coupled racetrack resonator with X-cut PPLN in one arm of the racetrack. The double resonance condition is achieved by thermal tuning of the MZI and the resonator by placing resistive heaters (e.g. NiCr as marked by yellow in Fig.\,\ref{fig:Fig1}(a)) on top of the waveguides with an oxide cladding layer in between. With a continuous pump in mode $b$ in the sub-threshold regime (i.e. below the lasing threshold of the OPO), vacuum mode $a$ in the resonator cavity evolves under the Hamiltonian of Eq.\,\ref{Eq_H} to generate the squeezed vacuum state. For an OPO satisfying the doubly resonant condition, the squeezing and anti-squeezing of the vacuum mode $a$ as a function of the pump power $P_\mathrm{s}$ in mode $b$, can be expressed as (Eq.\,(S49) in the supplementary section S5\,\cite{supp}),

\begin{equation}\label{squeezing1}
    S_\pm = 1 \pm \eta_\mathrm{a}\eta_\mathrm{T}\eta_\mathrm{d} \frac{4 \sqrt{\frac{P_\mathrm{s}}{P_\mathrm{th}}}}{ (1\mp  \sqrt{\frac{P_\mathrm{s}}{P_\mathrm{th}}})^2 + 4(1-\eta_\mathrm{a})^2(\frac{\Delta\omega_\mathrm{LO}}{\kappa_{a,0}})^2}
\end{equation}

where, $\Delta\omega_\mathrm{LO}$ is the local oscillator frequency, $\eta_\mathrm{T}$ is the transmittance of the squeezed state from chip to detector, and $\eta_\mathrm{d}$ is the quantum efficiency of the detector. In our calculation, we assume $\eta_\mathrm{T}$ and $\eta_\mathrm{D}$ to be unity. High-efficiency fiber-to-chip couplers have already been demonstrated on the TFLN platform \cite{he2019low,hu2021high,chen2022low}. On-chip quantum limited optical parametric amplifications on the same TFLN platform can also be used to overcome the limitations of imperfect transmission from chip to detector \cite{nehra2022few}. However, even if $\eta_\mathrm{T}$ and $\eta_\mathrm{D}$ reach unity, the maximum squeezing level that can be obtained from an OPO is fundamentally limited by the coupling ratio of mode $a$ no matter how efficient the internal generation process is \cite{park2024single}. Hence, the primary objective of this RMZI-PPLN design is to achieve high over-coupling conditions for attaining a higher level of squeezing with a relatively low pump threshold.

The theoretical pump power threshold $P_\mathrm{th}$ for the OPO is given by (Eq.\,(S50) in the supplementary section S5\,\cite{supp}),
\begin{equation}\label{squeezing2}
    P_\mathrm{th} = \left(1 + \frac{\delta^2 (1-\eta_a)^2}{\kappa_{a,0}^2}\right)\frac{\hbar\omega_b}{64g_0^2} \frac{\kappa_{a,0}^2}{(1-\eta_a)^2} \frac{\kappa_{b,0}}{\eta_b(1-\eta_b)} 
\end{equation}

Here, $\kappa_{i,\mathrm{0}}$ is the intrinsic loss rate for mode $i = \mathrm{\{a, b\}}$. Key factors such as the squeezing level, bandwidth, and threshold power characterize the efficiency of the generation process for the squeezed light. The external coupling rate strongly impacts both the squeezing level and bandwidth. From Eq.\,(\ref{squeezing1}), overcoupling mode $a$ directly enhances the squeezing (and anti-squeezing) level. Furthermore, a stronger over-coupling condition ensures a larger cavity bandwidth, thereby increasing the bandwidth of the squeezing and anti-squeezing light. However, for generating squeezing light efficiently, low $P_\mathrm{th}$ is also critical. A low $P_\mathrm{th}$ is necessary to minimize thermal effects and reduce the noise of the pump laser \cite{dutt2016tunable}. An ultra-low $P_\mathrm{th}$ may enable a chip-level squeezed light source where on-chip lasers can be used for the pump laser \cite{li2022integrated}. $P_\mathrm{th}$ depends on the coupling ratios of both modes $a$ and $b$ as per Eq.\,(\ref{squeezing2}). For a constant $\eta_\mathrm{a}$, $P_\mathrm{th}$ is minimized when $\eta_\mathrm{b}\left(1-\eta_\mathrm{b}\right)$ is maximized. This occurs when $\eta_\mathrm{b} = 0.5$, indicating critical coupling of the pump mode. On the contrary, $P_\mathrm{th}$ increases with increasing overcoupling condition of mode $a$. Consequently, simultaneously achieving high squeezing levels and low threshold power is challenging. Our proposed RMZI-PPLN device can address this trade-off in a controlled manner not relying on strict fabrication, enabling the optimization of the coupling conditions of the modes actively. Through thermal tuning of both the ring and the MZI, a highly over-coupled mode $a$ and a critically coupled mode $b$ can be achieved simultaneously for efficient generation of squeezed light. Fig.\,\ref{fig:Fig2}(d) shows that when mode $b$ is critically coupled, four distinct over-coupling ratios for mode $a$ can be achieved by adjusting the base temperature of the MZI within the range of 5\,K to 40\,K. The values of these coupling ratios and corresponding $\Delta T_\mathrm{MZI}$ are as following: $\eta_\mathrm{a\left(A\right)}\simeq0.98\,$ at $\Delta T_\mathrm{MZI} = 12.85\,\mathrm{K}$, $\eta_\mathrm{a\left(B\right)}\simeq0.98\,$ at $\Delta T_\mathrm{MZI} = 15.85\,\mathrm{K}$, $\eta_\mathrm{a\left(C\right)}\simeq0.88\,\mathrm{K}$ at $\Delta T_\mathrm{MZI} = 39.45\,\mathrm{K}$ and $\eta_\mathrm{a\left(D\right)}\simeq0.94\,$ at $\Delta T_\mathrm{MZI} = 42.16\,\mathrm{K}$. 
To achieve both low $P_\mathrm{th}$ and high level of squeezing, $\eta_\mathrm{a(D)}\simeq0.94\,$ is selected. 
Under the chosen coupling conditions for the mode $a$ and mode $b$, the anti-squeezing and squeezing levels in dB as a function of the input pump power for varying $\Delta\omega_\mathrm{LO}$ are shown in Fig.\,\ref{fig:Fig4}(a). The proposed scheme can achieve $\sim-12\,$\,dB squeezing level and $\sim30$\,dB anti-squeezing level with a sub-milliwatt pump power of -2.2\,dBm ($\sim610 \mathrm{ \,\mu W}$) when $\Delta\omega_\mathrm{LO} = 50$\,MHz. 
The squeezing level can be further improved by choosing a higher overcoupling condition if the $P_\mathrm{th}$ requirement is made less stringent. The anti-squeezing and squeezing levels are depicted in Fig.\,(S2) of the supplementary for different aforementioned attainable $\eta_\mathrm{a}$s (Point $A, B, C$ and $D$ in Fig.\,\ref{fig:Fig2}(d)). Maximum squeezing levels of -19\,dB, -17.6\,dB, -9\,dB, and -12\,dB are achievable with $P_\mathrm{th}$ values of $15$\,mW, $7.7$\,mW, $154\,\mu$W and $610\,\mu$W, respectively. The corresponding squeezing bandwidths are $7.86$\,GHz, $5.57$\,GHz, $0.70$\,GHz, and $1.5$\,GHz, respectively. Fig.\,\ref{fig:Fig4}(b) shows that the neighboring modes around the energy-matched resonant mode $a$ have significantly higher threshold power than that of the on-resonant one. This enables strong suppression of sideband noise in the generated squeezed light. 

\section{Parametric down conversion}
Apart from the generation of squeezed light, TFLN PPLN waveguides and ring resonators are suitable for generating high-efficiency single photons required for quantum computation and quantum information processing \cite{xin2022spectrally}. In this section, we show in detail how the proposed device can be used to generate single photons with a high purity and high heralding efficiency simultaneously. Furthermore, we propose and optimize a dual pulse pump scheme to increase the single photon purity without sacrificing high heralding efficiency. For both LOQC \cite{knill2001scheme} or FBQC \cite{bartolucci2023fusion}, integrated heralded single-photon sources must manifest the following attributes: (i) high heralding efficiency leading to a very low probability of vacuum amplitude in the heralded state; (ii) spectrally pure and unentangled heralded state (iii) high SPDC conversion efficiency with a high heralding rate and bandwidth (iv) and very low probability of heralding multi-photon states by low pump power operation. Several trade-offs among these competing requirements make the design of SPDC source very stringent on fabrication tolerance. Tunability of the conversion process can offer judicial control over the whole design space.
\begin{figure}[b]
    \centering
    \includegraphics[width = 0.45\textwidth]{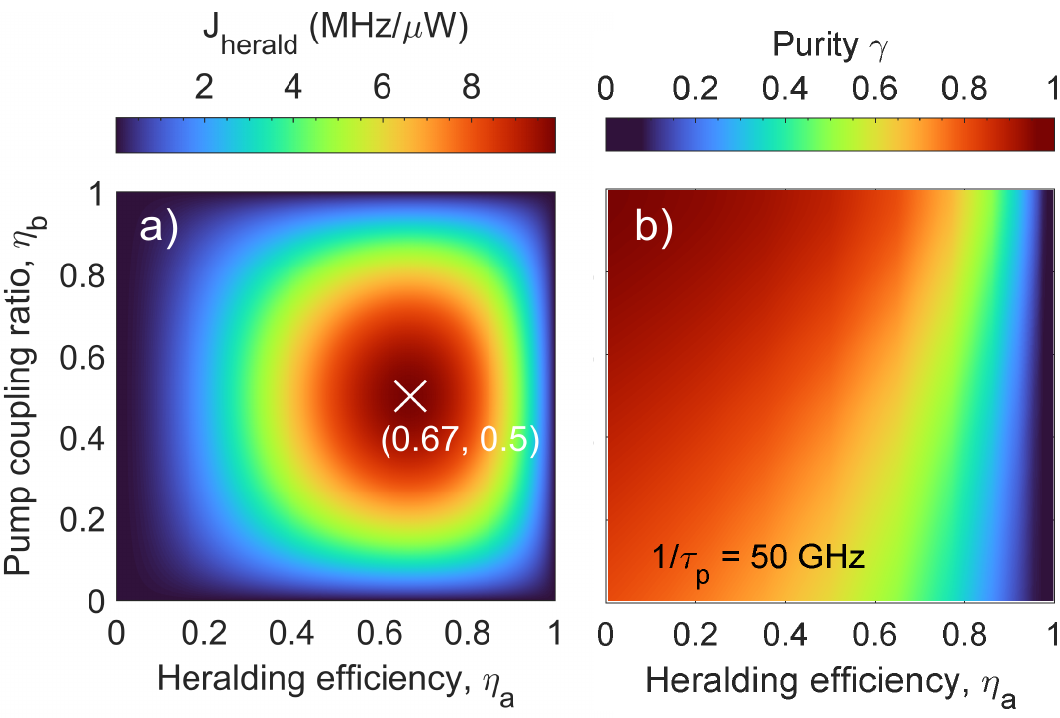}
    \caption{With intrinsic loss rate $\kappa_{a,0}/\mathrm{2\pi} = 98.5\,\mathrm{MHz}$ and $\kappa_{b,0}/\mathrm{2\pi} = 93.5\,\mathrm{MHz}$, pump and signal coupling ratios are varied to calculate, a) Successful herald rate, $J_\mathrm{herald}$ for continuous pumping scheme, the cross denotes the case at maximum herald rate b) Spectral purity, $\gamma$ for single pulse scheme with pulse width $\tau_\mathrm{p} = 0.02\,\mathrm{ns}$.}
    \label{fig:Fig5}
\end{figure}

Under the energy and momentum conservation of photonic modes, the non-linear three-wave mixing process can be described by the Hamiltonian in Eq.\,\ref{Eq_H}. A pump photon in mode $b$ parametrically down-converts into a pair of degenerate signal and idler photons in mode $a$, where the signal photon heralds the existence of a single idler photon in the appropriate channel.
Heralding efficiency of the generated photon at the telecom band is defined as $\eta_a = \frac{\kappa_{a,\mathrm{ex}}}{\kappa_{a,\mathrm{tot}}}$. A highly over-coupled resonator in the telecom wavelength is required to achieve high heralding efficiency. It should be noted that both coupling ratio and heralding efficiency for mode $a$ are denoted by $\eta_\mathrm{a}$ as they are identical parameters but have different names due to nomenclature conventions in different contexts. With the proposed thermally tuned RMZI-PPLN, such a highly over-coupled resonator with broad linewidth can be achieved, as shown in Fig.\,\ref{fig:Fig2}(c).
\begin{figure*}[!ht]
        \centering
    \includegraphics[width = 1.05\textwidth]{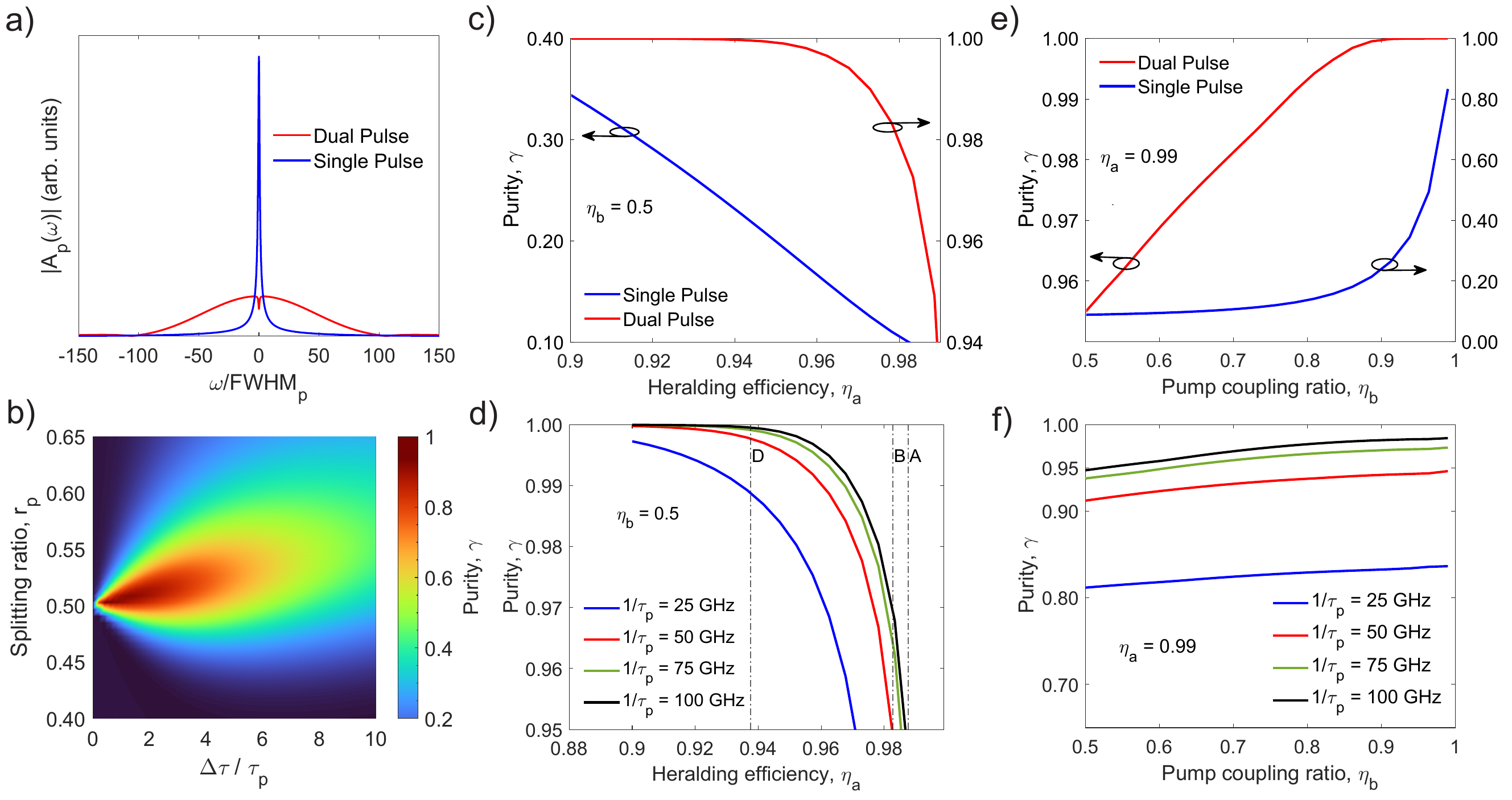}
    \caption{For the case of pulse width $\tau_p = 0.02$\,ns and pump resonance linewidth $\mathrm{FWHM_p} = 0.19\,\mathrm{GHz}$ corresponding to the MZI tuning point A in Fig.\,\ref{fig:Fig2}(d), a) Spectral profile of the in-resonator pump field for both single- and dual-pulse configurations. The parameters for the dual pulse is set to $\Delta\tau/\tau_p = 2.5$ and $r_p = 0.51$. b) Purity $\gamma$, as a function of the dual-pulse splitting ratio, $r_p$ and temporal separation, $\Delta\tau$. With sufficiently broad pump pulse to excite the full pump resonance, maximum achievable spectral purity is plotted, c) for pump mode being critically coupled and  the telecom mode varying around maximally overcoupled conditions and e) for telecom mode being highly overcoupled and pump mode's coupling ratio gradually increasing. Imposing different practical pulse widths, same cases are plotted in d) and f) for dual pulse scheme, showing slight deviations from optimum purity as pump pulse bandwidth drops. Vertical lines in d) denote different coupling conditions that were manifested by MZI tuning in Fig.\,\ref{fig:Fig2}(d). }
    \label{fig:Fig6_7}
\end{figure*}
However, the trade-off between heralding rate and heralding efficiency prevents achieving the maximum possible heralding rate at this highly over-coupled condition at the telecom wavelength, as evident from Fig.\,\ref{fig:Fig5}(a). Details of the derivation for the heralding rate can be found in the supplementary section S4\,\cite{supp}. $J_\mathrm{herald}$ is maximized at $\eta_a = 0.67$ and $\eta_b = 0.5$, but at our operative coupling conditions (operating point A denoted in Fig.\,\ref{fig:Fig2}(d), successful heralding rate $J_\mathrm{herald}$ is  $0.81\,\mathrm{ MHz/\mu W}$ with a bandwidth of $\kappa_a/2\pi = 7.92\,\mathrm{ GHz}$. This large heralding rate with low power is a direct attribute of the high nonlinear coefficient of TFLN platform compared to other $\chi^{(2)}$ materials such as AlN. 

High spectro-temporal purity is another essential attribute for the heralded photons since low purity degrades the fidelity of subsequent operations in a scalable quantum photonic platform \cite{mosley2008heralded}, and a specific pumping scheme is a crucial factor in ensuring the spectral quality of the generated photon pairs. Achieving high heralding efficiency along with high spectral purity demands sophisticated femtosecond pump pulses \cite{weston2016efficient} or the engineering of the crystal nonlinearity \cite{branczyk2011engineered}. However, as we will show, our proposed device can obtain high purity while preserving the heralding efficiency using only pico-second (ps) pump pulses. Continuous pump scheme, by its inherent nature, has pump bandwidth orders of magnitude narrower than output bandwidth $\kappa_a/2\pi$ employed, which intrinsically introduces spectral correlation in the bi-photon Joint Spectral Amplitude (JSA) of the generated photon pairs and consequently leads to very low spectral purity \cite{helt2010spontaneous}. The straightforward approach to make the JSA factorizable and to achieve arbitrarily high ($>99\,\%$) spectro-temporal purity is to broaden the effective in-resonator pump resonance linewidth, $A_\mathrm{p}(\omega) = \alpha_\mathrm{p}(\omega)l_\mathrm{p}(\omega)$, compared to the signal-idler resonances linewidth, where $\alpha_\mathrm{p}(\omega)$ and $l_\mathrm{p}(\omega)$ are respectively input pump pulse spectrum and resonance line shape function. For a single-pulse pumping scheme with pump pulse bandwidth of around $50\,\mathrm{ GHz}$, heralded photon spectral purity has been calculated and the result is shown in Fig.\,\ref{fig:Fig5}(b) (details of the derivation can be found in the supplementary section S4\,\cite{supp}). The upper portion of the figure with over-coupled signal and wide pump resonance line-shape, $l_\mathrm{p}(\omega)$ shows a region of high spectral purity. However, even if we impose an arbitrary broad input pump bandwidth, the purity will still be limited by an upper bound. Furthermore, this region directly contradicts the heralding efficiency, rate optimization, and low pump power requirements. Even if we further employ an arbitrarily broad spectrum, purity will still be bounded by an upper limit since $\alpha_\mathrm{p}(\omega)l_\mathrm{p}(\omega)$ is spectrally limited by the resonance linewidth. To overcome this trade-off to achieve arbitrarily high purity while preserving high heralding efficiency and low pump power, we analyze the purity of the generated photons when the device is pumped with a dual-pulse scheme. Both the single-pulse and dual-pulse spectra are shown in Fig.\,\ref{fig:Fig6_7}(a). Dual pulses are characterized by the pulse splitting ratio $r_p$ and the temporal separation between the pulses $\Delta\tau$ and the pulse width $\tau_p$ as described in detail in the supplementary section S4\,\cite{supp}. The dual-pulse scheme provides a much broader in-resonator pump spectrum than the corresponding resonance linewidth as shown in Fig.\,\ref{fig:Fig6_7}(a). In Fig.\,\ref{fig:Fig6_7}(b), we show the spectral purity where the dual-pulse parameters are varied with a fixed pump pulse width of $0.02\mathrm{\,ns}$ and the linewidth parameters for the mode a and b are set by the operating point A as shown in Fig.\,\ref{fig:Fig2}(d), which demonstrates the availability of a large parameter space capable of providing high spectral purity. In Fig.\,\ref{fig:Fig6_7}(c) and Fig.\,\ref{fig:Fig6_7}(e), we show the maximum achievable purity for different heralding efficiencies and coupling ratios for the pump mode b. For both cases, a sufficiently broad pump pulse spectrum is chosen to achieve the upper bound of the purity, and corresponding optimized dual-pulse parameters have been used for the dual-pulse scheme. In contrast to Fig.\,\ref{fig:Fig5}(b), the dual-pulse scheme shifts the upper limit of purity arbitrarily close to unity even with high heralding efficiency along with a critically coupled pump mode. Single photon purity for optimized dual-pulses having different pump pulse bandwidths is shown in Fig.\,\ref{fig:Fig6_7}(d) and Fig.\,\ref{fig:Fig6_7}(f). It is evident that purity tends to decrease significantly if pump bandwidth drops below $50\mathrm{\,GHz}$. Nevertheless, it is important to note that intrinsic losses associated with the telecom mode and near-IR pump mode are crucial factors for the purity of the generated single photons. The purity of the single photons can further be improved if the intrinsic losses are reduced. For instance, for a case of 99\,\% heralding efficiency with critically coupled $50\mathrm{\,GHz}$ dual-pulse pump, purity can be enhanced by 6\,\% if the losses can be reduced by 2\,times.

\begin{figure}[t]
        \centering
    \includegraphics[width = 0.35\textwidth]{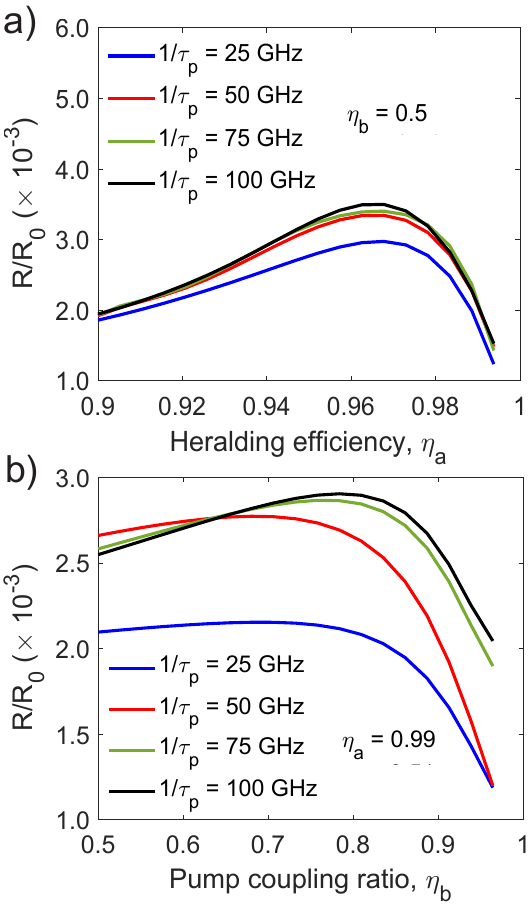}
    \caption{For different pump pulse bandwidths, the dual-pulse scheme is optimized for maximum purity, and corresponding relative pair generation probability $R/R_0$ is plotted a) varying coupling condition of the telecom mode, b) varying coupling condition of the pump mode. Here, $R_0$ is the generation rate in the single-pump scheme with corresponding pump pulse bandwidth.}
    \label{fig:Fig8}
\end{figure}
Though the dual-pulse scheme facilitates a larger in-resonator bandwidth of the pump field and improves the photon purity, however, it also significantly reduces the peak pump intensity, eventually leading to the degradation of the generation efficiency. In Fig.\,\ref{fig:Fig8}, normalized photon generation rate relative to the single-pump scheme is plotted for the optimized parameters for high purity (as shown in Fig.\,\ref{fig:Fig6_7}(d) and Fig.\,\ref{fig:Fig6_7}(f)). Even though the generation rates degrade to almost three orders of magnitude lower than the single-pulse case, this can be improved by proportionately increasing the input peak power of the pump pulse. Since the three-wave-mixing interaction for the SPDC process has a direct proportional relationship between the generation probability $R$ and pump pulse energy $E_p$ ($R\propto E_p$), similar generation rates can be achieved as shown in Fig.\,\ref{fig:Fig5}(a) by increasing the effective pump pulse energy. For a pump source of repetition rate $50\mathrm{\,MHz}$ and pulse duration $0.02\mathrm{\,ns}$, a similar generation rate will require a $1\mathrm{\,W}$ of peak pump power, which can be achievable with the table-top mode-locked lasers. Especially recently demonstrated on-chip mode lock lasers on the TFLN platform \cite{guo2023ultrafast} can be an ideal source for the pump pulse offering fully compact on-chip solutions for high purity high heralding efficiency single photon source.
\section{Conclusion} 
In summary, an interferometrically coupled ring resonator on the TFLN platform with tunable coupling for the telecom and near IR modes has been proposed. By adjusting the temperatures of the MZI and the ring using thermal heaters, the cavity transmission spectrum and the resonant wavelengths can be actively tuned to achieve optimal coupling conditions. Such devices can be used to generate efficient squeezing light and single-photons in the telecom band $(\sim1550\mathrm{\,nm})$ with a pump mode in the near IR band $(\sim775\mathrm{\,nm})$. With the proposed device and scheme, on-chip $-12$\,dB squeezed light can be generated at a sub-milliwatt power of $610\,\mu$W having a bandwidth of 1.5\,GHz. Further improvements in the squeezing levels, reaching up to -19\,dB with a bandwidth of 7.86\,GHz, can be achieved at the cost of increasing pump power to 15\,mW. Moreover, single photons with simultaneous high heralding efficiency and high purity can be generated using the device employing a dual-pulsed pump. The dual-pulse scheme with parameters can provide close to unity purity for the generated single photons. Though the photon generation rate of the dual-pulse method is low compared to the single-pulse scheme, this can be mitigated by increasing the pump energy. Thus, the proposed device presents a generalized solution for generating highly squeezed light with high bandwidth at a low pump power, and high purity single photons with high heralding efficiency, both of which are crucial for scalable photonic quantum computation and quantum communication.    
\bibliography{Reference}
\onecolumngrid
\newpage
\setcounter{section}{0}
\setcounter{equation}{0}
\setcounter{figure}{0}
\renewcommand{\theequation}{S\arabic{equation}}
\renewcommand{\thefigure}{S\arabic{figure}} 

\section*{\textnormal{Supplementary Information for \\ Periodically poled thin-film lithium niobate ring Mach Zehnder coupling interferometer as an efficient quantum source of light}}



\section{Device and Simulation Details}
\begin{figure}[H]
    \centering
    \includegraphics[width = 1\textwidth]{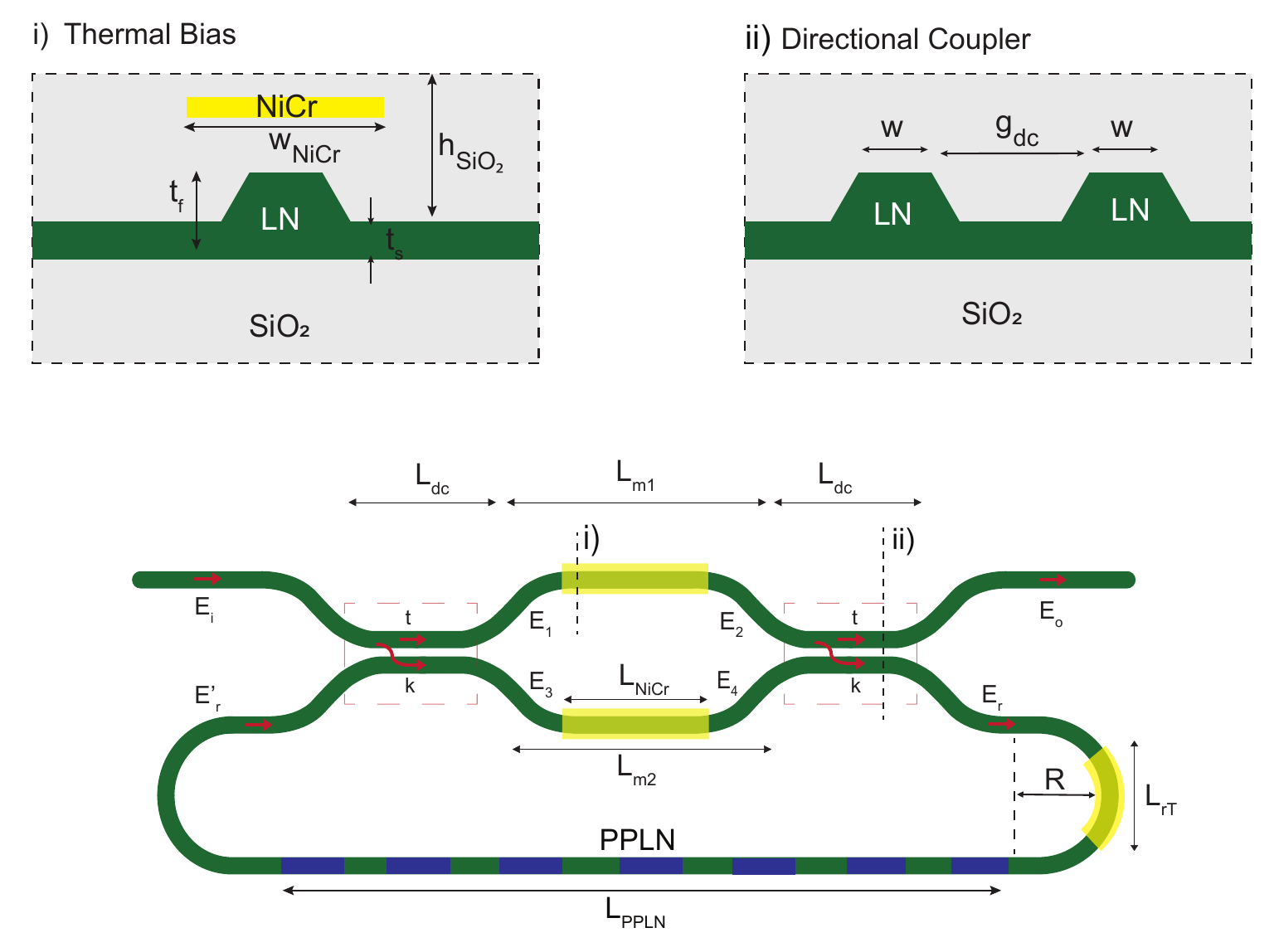}
    \caption{Simplified schematic of MZI coupler with the PPLN ring. i) the device-cross section at the location of the thermal heater (NiCr) and ii) the device-cross section of the directional coupler segment of the resonator. Different geometric parameters (L\textsubscript{dc}, g\textsubscript{dc}, w, h\textsubscript{SiO\textsubscript{2}}, L\textsubscript{PPLN} and R)  used in the simulation have been depicted.}
    \label{fig:RMZI2}
\end{figure}
We considered an  X-cut LN waveguide with $1.2 \mathrm{\mu m}$ width and $200 \mathrm{nm}$ etching height on a $200 \mathrm{nm}$ film thickness for both the directional coupler and resonator region. We chose $L_\mathrm{dc} = 398\mu\mathrm{m}$ and $g_\mathrm{dc} = 600\mu\mathrm{m}$ for the rest of the paper to achieve approximately 50\% and 10\% cross-transmission in the fundamental and second-harmonic mode respectively. Other dimensions were taken to be $L_\mathrm{m1} = L_\mathrm{m2} = 500\mu\mathrm{m}$ and $R = 80\mu\mathrm{m}$. Material dispersion and propagation loss for each of the fundamental TE modes are included into $\beta = \frac{2\pi}{\lambda}n_\text{eff}$ and $\alpha$. For telecom and near IR mode, propagation losses were both taken to be $0.2 \mathrm{dB/cm}$ which are comparable to the reported values \citeS{Sshams2022reduced,Sdesiatov2019ultra}.

\section{RMZI Characteristics}

The directional coupler (DC) is characterized by the through transmission and cross-transmission coefficients $t$ and $k$ respectively, and the transfer matrix can be written compactly as,
\begin{equation}
\begin{split}
    M_\mathrm{dc} & = \begin{pmatrix}
        t & ik \\ ik & t
    \end{pmatrix} \\
    t &= \sqrt{T} e^{-\alpha L_{\text{dc}}/2} e^{-i\beta L_{\text{dc}}} \\
    k &= \sqrt{1-T} e^{-\alpha L_{\text{dc}}/2} e^{-i\beta L_{\text{dc}}},\\
\end{split}
\end{equation}
    
where $T$ is the power transmission in the DC with a length of $L_\mathrm{dc}$ and gap of $g_\mathrm{dc}$. 


Defining phase terms in the ring arm, the upper and lower MZI arms with respective lengths of  $L_\mathrm{r}$, $L_\mathrm{m1}$ and $L_\mathrm{m2}$, 
\begin{equation}
\begin{split}
    \sigma_r & = e^{-\alpha L_\mathrm{r}/2} e^{-i\beta L_\mathrm{r}}\\
    \sigma_\mathrm{m1} & = e^{-\alpha L_\mathrm{m1}/2} e^{-i\beta L_\mathrm{m1}} \\
    \sigma_\mathrm{m2} & = e^{-\alpha L_\mathrm{m2}/2} e^{-i\beta L_\mathrm{m2}} 
\end{split}
\end{equation}
Transfer characteristics of an isolated MZI coupler without the ring feedback can be expressed as,
\begin{equation}
    \begin{split}
        \begin{pmatrix}
            E_\mathrm{o,up} \\ E_\mathrm{o,down}
        \end{pmatrix} 
        & = \begin{pmatrix}
            t & ik \\ ik & t
        \end{pmatrix}
        \begin{pmatrix}
            \sigma_\mathrm{m1} & 0 \\ 0 & \sigma_\mathrm{m2}
        \end{pmatrix}
         \begin{pmatrix}
            t & ik \\ ik & t
        \end{pmatrix}
        \begin{pmatrix}
            E_i \\ 0
        \end{pmatrix} \\
    \end{split}
\end{equation}

Comparing the whole MZI coupler to a point coupler and conforming to the definition in Ref. \citeS{Sbogaerts2012silicon}, we define the self-coupling coefficient $t_\text{m}$ and cross-coupling coefficient $k_\text{m}$ of the MZI, 

\begin{equation}
\label{SCC}
\begin{split}
    t_{\text{m}} =  \frac{E_{\text{o,up}}}{E_i} &= t^2 \sigma_\mathrm{m1} - k^2 \sigma_\mathrm{m2} \\  
    k_{\text{m}} =  \frac{E_{\text{o,down}}}{E_i} &= ikt(\sigma_\mathrm{m1} + \sigma_\mathrm{m2}) \\
\end{split}
\end{equation}

If the resonator has a round trip length of $L = L_\mathrm{r} + L_\mathrm{m2} + 2L_\mathrm{dc}$, then single-pass amplitude transmission, including both propagation loss in the ring and loss in the couplers, is defined as $a = e^{-\alpha L/2}$. Relative value between the self-coupling coefficient and the single-pass amplitude transmission indicates different coupling conditions, critical coupling is acheived when single pass amplitude exactly balances the self-coupling transmission, $|t_{\text{m}}| =  a$, and overcoupling and under coupling condition are achieved respectively for $|t_{\text{m}}| <  a$ and $|t_{\text{m}}| >  a$.

Transfer characteristics of the MZI coupler with the ring feedback can be expressed as,
\begin{equation}
    \begin{split}
        \begin{pmatrix}
            E_\mathrm{o} \\ E_\mathrm{r}
        \end{pmatrix}
        & = \begin{pmatrix}
            M_{11} & M_{12}  \\ M_{21}  & M_{22} 
        \end{pmatrix}  
        \begin{pmatrix}
            E_\mathrm{i} \\  E_\mathrm{r}
        \end{pmatrix} \\
        & = \begin{pmatrix}
            t & ik \\ ik & t
        \end{pmatrix}
        \begin{pmatrix}
            \sigma_\mathrm{m1} & 0 \\ 0 & \sigma_\mathrm{m2}
        \end{pmatrix}
         \begin{pmatrix}
            t & ik \\ ik & t
        \end{pmatrix}
        \begin{pmatrix}
            1 & 0 \\ 0 & \sigma_r
        \end{pmatrix}
        \begin{pmatrix}
            E_i \\ E_\mathrm{r}
        \end{pmatrix} \\
    \end{split}
\end{equation}

Here, 
\begin{equation*}
    \begin{split}
        E_\mathrm{r} &= \frac{M_{21}}{1-M_{22}} E_\mathrm{i} \\
        E_\mathrm{o} &= \left[ M_{11} + \frac{M_{21}}{1-M_{22}} \right]E_\mathrm{i}
                    = \frac{M_{11} - Det(M)}{1-M_{22}}E_\mathrm{i}
    \end{split}
\end{equation*}

where, 
\begin{equation*}
    \begin{split}
        M_{11} &= t^2\sigma_\mathrm{m1} - k^2\sigma_\mathrm{m2}\\
        M_{22} &= \sigma_r(t^2\sigma_\mathrm{m2} - k^2\sigma_\mathrm{m1})\\
        Det(M) &= \sigma_r \sigma_\mathrm{m1} \sigma_\mathrm{m2} (t^2 + k^2)^2
    \end{split}
\end{equation*}
Finally cavity transmission is expressed by, 
\begin{equation} 
\label{Pout}
\begin{split}
    P_\mathrm{out} &= \left|\frac{E_o}{E_i}\right|^2=\left|\frac{\left(t^2\sigma_\mathrm{m1}-k^2\sigma_\mathrm{m2}\right)-\sigma_r \sigma_\mathrm{m1}\sigma_\mathrm{m2}(t^2+k^2)^2}{1-\sigma_r(t^2\sigma_\mathrm{m2}-k^2\sigma_\mathrm{m1})}\right|^2. 
\end{split}
\end{equation}
For a lossless point coupler, i.e. $ t = \sqrt{T}$, $ k = \sqrt{1-T}$ and $|t|^2+|k|^2=1$ and  symmetric MZI, i.e. $L_\mathrm{m}=L_\mathrm{m1} = L_\mathrm{m1}$, $\sigma = \sigma_\mathrm{m1} = \sigma_\mathrm{m2}$ ,
\begin{equation} 
\begin{split}
    P_\mathrm{out} = \left|\frac{E_o}{E_i}\right|^2
    &= \left|\frac{t_m-\sigma^2\sigma_r}{1-\sigma^2\sigma_rt_m^*}\right|^2. \\
\end{split}
\end{equation}

\section{Thermal Tuning Calculation}
Our designed scheme of control incorporates two independent NiCr resistive thermal tuners, one tuner at the MZI arms in push-pull configuration, and another single tuner at the PPLN ring arm. By changing the relative phase shift between the bus and ring, MZI tuner directly modifies the total coupling rates of the telecom and near IR mode. To have relatively flat coupling conditions around each of the fundamental modes, we opt for symmetric MZI ($L_\mathrm{m1} = L_\mathrm{m1} =L_\mathrm{m}$) with an effective MZI thermal tuner length of $L_{\text{mT}} = 0.5 L_\text{m}$. On the other hand, tunner in the resonator arm is used for fine control on resonance wavelength to satisfy the frequency matching condition between the fundamental modes. Effective length in the ring thermal tuner is chosen to $L_{\text{rT}} = 0.2 L_\text{r}$. Using Eq. \ref{Pout}, final cavity transmission for the symmetric RMZI can be found by incorporating the thermal phase shifts,

\begin{equation} 
\begin{split}
    \sigma_\mathrm{m1} &= e^{-\alpha L_\mathrm{m}/2} e^{-i(\beta L_\mathrm{m} + \phi_\text{MZI})}\\
    \sigma_\mathrm{m2} &= e^{-\alpha L_\mathrm{m}/2} e^{-i(\beta L_\mathrm{m} - \phi_\text{MZI})}\\
    \sigma_r &= e^{-\alpha L_\mathrm{r}/2} e^{-i(\beta L_\mathrm{r} + \phi_\text{ring})}
\end{split}
\end{equation}
Because of the difference in thermo-optic coefficient $\frac{dn_\text{{eff}}}{dT}$ of TFLN waveguide for 1550nm and 775nm \citeS{Sliu2022thermally}, amount of phase shift in the MZI arms will be different for both modes. Phase change in the lower MZI arm for a change in push-pull MZI temperature $\Delta T_\text{MZI}$
\begin{equation}
\begin{split}
    \phi_{\text{MZI}} &= \left(\frac{2\pi}{\lambda} \Delta n_{\text{eff}}\right) L_{\text{mT}}
\end{split}
\end{equation}
Where, 
\begin{equation*}
    \Delta n_{\text{eff}} = \int_{T_0}^{T_{\text{MZI}}} \frac{dn_\text{eff}}{dT} dT \sim \Delta T_{\text{MZI}}\frac{dn_\text{eff}}{dT}
\end{equation*}
Likewise, phase change in the resonator arm for change in ring temperature $\Delta T_\text{ring}$,
\begin{equation}
\begin{split}
   \phi_{\text{ring}} &= \left(\frac{2\pi}{\lambda} \Delta n_{\text{eff}}\right) L_{\text{rT}}
\end{split}
\end{equation}


From Eq. \ref{SCC}, self-coupling coefficient can easily shown to be, 
\begin{equation}
\label{tm}
\begin{split}
    |t_m| &= e^{-\alpha L_m/2}e^{-\alpha L_\mathrm{dc}} \sqrt{(2T-1)^2 \cos^2{(\phi_\mathrm{MZI})} + \sin^2{(\phi_\mathrm{MZI})}}\\
    \mathrm{arg} (t_m) &= - \beta (L_m + 2L_\mathrm{dc}) - \arctan \left[\frac{\tan{(\phi_\mathrm{MZI})}}{2T-1}\right]
\end{split}
\end{equation}
It is evident from Eq. \ref{tm} that for a 50-50 directional coupler, introduction of thermal phase shift in the MZI arm will not affect the phase of the self-coupling coefficient, thus, resonance peak of the overall cavity transmission will not shift.




 
However, shift in the cavity resonance in response to the change in ring temperature can be found,
\begin{equation*}
\begin{split}
    \frac{2\pi}{\lambda'} n_\text{neff} (L_\mathrm{r} +2L_\mathrm{dc} + L_\mathrm{m}) + \frac{2\pi}{\lambda'} \Delta n_\text{neff} L_\text{rT}  &=   \frac{2\pi}{\lambda} n_\text{neff} (L_\mathrm{r} +2L_\mathrm{dc} + L_\mathrm{m}) \\
    \Delta \lambda &= \Delta T_\text{ring} \frac{\lambda}{n_\text{neff}}  \frac{L_\text{rT} }{(L_\mathrm{r} +2L_\mathrm{dc} + L_\mathrm{m})} \frac{dn_\text{neff}}{dT}
\end{split}
\end{equation*}

For the ring arm thermal tuner of our particular device, resonance shift in the telecom and near IR mode are calculated to be $\Delta\lambda = 1.22\mathrm{pm/K}$ and $\Delta\lambda = 0.72\mathrm{pm/K}$ respectively. If temperature resolution is $\delta T = 0.1\mathrm{K}$ then shift in frequency can be written as
\begin{equation*}
    \begin{split}
       \textbf{Telecom:}  \Delta f & = -\frac{c}{\lambda^2} \Delta \lambda  = 15 \mathrm{MHz} \\
       \textbf{Near IR:}   \Delta f & = -\frac{c}{\lambda^2} \Delta \lambda  = 36 \mathrm{MHz} \\
    \end{split}
\end{equation*}



\section{Degenerate SPDC in $\chi^{2}$ microring resonator}
The interaction Hamiltonian, $\mathcal{H}_\mathrm{I}$ is given by:
\begin{equation}\label{eq1}
    \mathcal{H}_\mathrm{I} = \hbar g_{eo}({a}+{a}^\dagger)^2({b}+{b}^\dagger)
\end{equation}
Here, ${a}$ and ${b}$ are the bosonic annihilation operators for visible and telecom modes, respectively. The frequencies of the modes are related by: $\omega_b = 2\omega_a$.
\begin{equation}\label{eq2}
\begin{split}
    \mathcal{H}_\mathrm{I} & = \hbar g_{eo}\left({a}^2+{a}^\dagger{a}+{a}{a}^\dagger+({a}^\dagger)^2\right)\left({b}+{b}^\dagger\right) \\
    &= \hbar g_{eo}\left({a}^2b+{a}^\dagger{a}b+{a}{a}^\dagger b+({a}^\dagger)^2b +{a}^2{b}^\dagger+{a}^\dagger{a}{b}^\dagger+{a}{a}^\dagger{b}^\dagger+({a}^\dagger)^2{b}^\dagger\right) \\
    &= \hbar g_{eo}\left({a}^2b^\dagger+({a}^\dagger)^2b \right)
\end{split}
\end{equation}

We have ignored the rapidly rotating terms. Now, the complete Hamiltonian is given by:
\begin{equation} \label{eq5}
\mathcal{H}=\hbar \omega_a {a}^{\dagger} {a}+\hbar \omega_b {b}^{\dagger} {b}+\mathcal{H}_{\mathrm{I}}+i \sqrt{\frac{\kappa_{b, \mathrm{ex}} P_p}{\hbar \omega_{\mathrm{p}}}}\left(b^{\dagger} e^{-i \omega_{\mathrm{p}} t}- b e^{i \omega_{\mathrm{p}} t}\right)
\end{equation}
Now, let's do a rotating frame transformation ${A} = \hbar\frac{\omega_p}{2} {a}^\dagger {a} + \hbar\omega_p {b}^\dagger {b}$, where $\omega_p$ is the pump frequency. Now, the corresponding unitary matrix is ${U} = e^{i{A}t/\hbar} = e^{i\omega_pt{a}^\dagger{a}/2}e^{i\omega_pt{b}^\dagger{b}}$. Now the transformed Hamiltonian $\mathcal{H}_t$ is given by:
\begin{equation}
    \mathcal{H}_t = i\hbar\dot{{U}}{U}^\dagger + {U}\mathcal{H}{U}^\dagger
\end{equation}
Now, $i\hbar\dot{{U}}{U}^\dagger = -\hbar\left(\frac{\omega_p}{2} {a}^\dagger {a} + \omega_p {b}^\dagger {b}\right) = -\hbar\frac{\omega_p}{2}{a}^\dagger{a} - \hbar\omega_p{b}^\dagger{b}$. \\

The first four terms of the Hamiltonian in eqn (\ref{eq5}) remain invariant under $
{U}\mathcal{H}{U}^\dagger$ transformation. \\
\begin{equation*}
\begin{split}
    {U}{b}e^{i\omega_pt}{U}^\dagger &= \left[e^{i\omega_pt{a}^\dagger{a}/2}e^{i\omega_pt{b}^\dagger{b}} \right]{b} e^{i\omega_pt} \left[e^{-i\omega_pt{a}^\dagger{a}/2}e^{-i\omega_pt{b}^\dagger{b}}\right] \\
    & = e^{i\omega_pt}\left({b}+i\omega_pt\left[{b}^\dagger{b},b\right]+\frac{(i\omega_pt)^2}{2!}\left[{b}^\dagger{b},\left[{b}^\dagger{b},b\right]\right]\cdots\right) \\
    & = e^{i\omega_pt}\left({b}-i\omega_pt{b}+\frac{1}{2!}(i\omega_pt)^2{b}\cdots\right) \\
    & = e^{i\omega_pt}{b}\left(1-i\omega_pt+\frac{1}{2!}(i\omega_pt)^2\cdots\right) \\
    & = e^{i\omega_pt}{b}e^{-i\omega_pt} \\
    & = {b}
\end{split}
\end{equation*}
Similarly, ${U}{b}^\dagger e^{-i\omega_pt}{U}^\dagger = {b}^\dagger$. While deriving this, we used Baker-Campbell-Housdorff Equation and $\left[{b}^\dagger{b}, {b}\right] = -{b}$. So the transformed Hamiltonian,

\begin{equation}\label{eq7}
\begin{split}
    \mathcal{H}_\mathrm{t} & = \hbar\left(\omega_a-\frac{\omega_p}{2}\right){a}^\dagger{a}+\hbar\left(\omega_b-\omega_p\right){b}^\dagger{b}+\mathcal{H}_\mathrm{I}+i \sqrt{\frac{\kappa_{b, \mathrm{ex}} P_\mathrm{p}}{\hbar \omega_{\mathrm{p}}}}\left(b^{\dagger} -b\right) \\
    & = \hbar\delta_a{a}^\dagger{a}+\hbar\delta_b{b}^\dagger{b}+\mathcal{H}_\mathrm{I}+i \sqrt{\frac{\kappa_{b, \mathrm{ex}} P_\mathrm{p}}{\hbar \omega_{\mathrm{p}}}}\left(b^{\dagger} -b\right)
\end{split}
\end{equation}

Here $\kappa_{b,\mathrm{ex}}$ is the external coupling rate  definition $Q_b = \omega_b/\kappa_{b,\mathrm{ex}}$, and $\delta_a = \omega_a-\frac{\omega_p}{2}$ and  $\delta_b = \omega_b-\omega_p$ is the detuning for mode $a$ and $b$. Now, we assumed mode ${b}$ to be a strong pump mode and non-depleted. So, ${b}$ mode is in steady state and $d{b}/dt = 0$. Now, 
\begin{equation}
    \begin{split}
        \frac{d}{dt} b & = \frac{i}{\hbar}\left[\mathcal{H}_\mathrm{t}, {b}\right]-\frac{\kappa_{b,\mathrm{tot}}}{2}{b} \\
        \Rightarrow & 0  = \frac{i}{\hbar}[\hbar\delta_b{b}^\dagger{b}, {b}]+i\cdot i\sqrt{\frac{\kappa_\mathrm{b,ex}P_\mathrm{p}}{\hbar\omega_p}} - \frac{\kappa_{b,\mathrm{tot}}}{2} b\\
        \Rightarrow & -i\delta_b{b}-\sqrt{\frac{\kappa_\mathrm{b,ex}P_\mathrm{p}}{\hbar\omega_p}}-\frac{\kappa_{b,\mathrm{tot}}}{2} b= 0 \\
        \Rightarrow & {b} = \sqrt{\frac{\kappa_\mathrm{b,ex}P_p}{\hbar\omega_p}}\frac{1}{\sqrt{\delta_b^2+\left(\frac{\kappa_{b,\mathrm{tot}}}{2}\right)^2}} e^{i\phi} \\
        \Rightarrow & {b} = \sqrt{\frac{\kappa_{b,\mathrm{ex}}}{\delta_a^2+\left(\frac{\kappa_{b,\mathrm{tot}}}{2}\right)^2}}\sqrt{\frac{P_p}{\hbar\omega_p}} e^{i\phi}
    \end{split}
\end{equation}
So, in the non-depletion and strong pump condition, we can assume pump mode $a$ to be a complex number. Also, for simplicity, $\phi = 0$ and so our previous assumption to model mode $b$ as a coherent state having a real $\beta$ is valid. So,
\begin{equation*}
      b = b^\dagger = \beta = \sqrt{\frac{\kappa_\mathrm{b,ex}}{\delta_b^2+\left(\frac{\kappa_{b,\mathrm{tot}}}{2}\right)^2}}\sqrt{\frac{P_p}{\hbar\omega_p}}
\end{equation*}
Now the hamiltonian becomes,
\begin{equation}
    \mathcal{H}_\mathrm{t} = \hbar\delta_b|\beta|^2+\hbar\delta_a{a}^\dagger{a}+\hbar g|\beta|\left[({a}^\dagger)^2+{a}^2\right]
\end{equation}

But the first term generates a trivial and constant time evolution. So we can ignore the first term. Therefore the final Hamiltonian becomes:
\begin{equation}
    \mathcal{H}_\mathrm{t} = \hbar\delta_a{a}^\dagger{a}+\hbar g|\beta|\left[({a}^\dagger)^2+{a}^2\right]
\end{equation}

For the SPDC, the Langevin dynamics of telecom photons with vacuum noise mode $a_\mathrm{in}$ is written as,

\begin{equation}
\begin{aligned}
     \frac{da}{dt} &=i[\mathcal{H},a] - \frac{\kappa_{a,\mathrm{tot}}}{2} a +\sqrt{\kappa_{a,\mathrm{tot}}} a_\mathrm{in}  \\
     &= \left[-i\delta_a - \frac{\kappa_{a,\mathrm{tot}}}{2}\right] a -i2g|\beta| a^\dagger +\sqrt{\kappa_{a,\mathrm{tot}}} a_\mathrm{in}  
\end{aligned}
\end{equation}
Taking Fourier transform operator $a(\omega) = \frac{1}{\sqrt{2\pi}} \int a(t) e^{i\omega t} dt$ on both side, equation of motion for mode $a(t)$ and its hermitian conjugate $a^\dagger(t)$ become,

\begin{equation}
    \begin{aligned}
    -i\omega a(\omega) &= \left[-i\delta_a -\frac{\kappa_{a,\mathrm{tot}}}{2} \right] a(\omega) - i2g|\beta| a^\dagger(-\omega) +\sqrt{\kappa_{a,\mathrm{tot}}} a_\mathrm{in} (\omega)\\
    i\omega a^\dagger(-\omega) &= \left[i\delta_a -\frac{\kappa_{a,\mathrm{tot}}}{2} \right] a^\dagger(-\omega) + i2g|\beta| a(\omega) +\sqrt{\kappa_{a,\mathrm{tot}}} a^\dagger_\mathrm{in} (-\omega)
    \end{aligned}
\end{equation}

\begin{equation}
\begin{aligned}
    a(\omega) & =\frac{- \sqrt{\kappa_{a, t o t}} \cdot\left[i\left(\delta_a+\omega\right)-\frac{\kappa_{a,\mathrm{tot}}}{2}\right] a_{i n}(\omega) -i2 g|\beta| \sqrt{\kappa_{a, t o t}} a_{i n}^{\dagger}(-\omega)}{\left[i\left(\delta_a+\omega\right)-\frac{\kappa_{a,\mathrm{tot}}}{2}\right] \cdot\left[-i\left(\delta_a-\omega\right)-\frac{\kappa_{a,\mathrm{tot}}}{2}\right]-4 g^2|\beta|^2} \\
    a^{\dagger}(-\omega) & =\frac{- \sqrt{\kappa_{a, t o t}} \cdot\left[-i\left(\delta_a-\omega\right)-\frac{\kappa_{a,\mathrm{tot}}}{2}\right] a_{i n}^{\dagger}(-\omega)+i2 g|\beta| \sqrt{ \kappa_{a, t o t}} a_{i n}(\omega)}{\left[-i\left(\delta_a-\omega\right)-\frac{\kappa_{a,\mathrm{tot}}}{2}\right] \cdot\left[i\left(\delta_a+\omega\right)-\frac{\kappa_{a,\mathrm{tot}}}{2}\right]-4 g^2|\beta|^2} .
\end{aligned}
\end{equation}

\subsection{Rate Calculation}
With the definition of $\langle a_\mathrm{in}(\omega) a_\mathrm{in}^\dagger(\omega^\prime) \rangle = \delta(\omega - \omega^\prime)$, photon number spectrum density is,
\begin{equation}
    \begin{aligned}
        \langle a^\dagger(\omega) a(\omega^\prime) \rangle &= \frac{ (2 g|\beta| \sqrt{\kappa_{a, t o t}})^2}{\left[-i\left(\delta_a+\omega\right)-\frac{\kappa_{a,\mathrm{tot}}}{2}\right] \cdot\left[i\left(\delta_a-\omega\right)-\frac{\kappa_{a,\mathrm{tot}}}{2}\right]} \\
        &\frac{\delta(\omega-\omega^\prime)}{\left[i\left(\delta_a+\omega^\prime\right)-\frac{\kappa_{a,\mathrm{tot}}}{2}\right] \cdot\left[-i\left(\delta_a-\omega^\prime\right)-\frac{\kappa_{a,\mathrm{tot}}}{2}\right] }
    \end{aligned}
\end{equation}

Total photon pair generation rate,
\begin{equation}
    \begin{aligned}
        R &= \frac{\kappa_{a,\mathrm{tot}}}{2} \langle a^\dagger(t) a(t^\prime) \rangle\\
        &=\frac{\kappa_{a,\mathrm{tot}}}{2} \frac{1}{2\pi}\int \int d\omega d\omega^\prime \langle a^\dagger(\omega) a(\omega^\prime) \rangle e^{i(\omega^\prime-\omega)t}\\
        &= \frac{\kappa_{a,\mathrm{tot}}}{2}\frac{1}{2\pi} \int\frac{ d\omega (2 g|\beta| \sqrt{ \kappa_{a, t o t}})^2}{\left[\left(\delta_a+\omega\right)^2+\left(\frac{\kappa_{a,\mathrm{tot}}}{2}\right)^2\right] \cdot \left[\left(\delta_a-\omega\right)^2+\left(\frac{\kappa_{a,\mathrm{tot}}}{2}\right)^2\right]}\\
        &= 2g^2 \frac{\frac{\kappa_{a,\mathrm{tot}}}{2}}{\delta_a^2 + \left(\frac{\kappa_{a,\mathrm{tot}}}{2}\right)^2} \frac{\kappa_{b,\mathrm{ex}}}{\delta_b^2 + \left(\frac{\kappa_{b,\mathrm{tot}}}{2}\right)^2} \frac{P_p}{\hbar\omega_p}
    \end{aligned}
\end{equation}
With zero detuning and non-depleted continuous pump power $P_p$ into visible mode $b$, the total photon pair generation rate in the telecom mode $a$ in the ring cavity is given by,\\
\begin{equation*}
    \begin{split}
    R &= 16g^2  \frac{\kappa_{b,\mathrm{ex}}}{\kappa_{b}^2} \frac{1}{\kappa_{a}} \frac{P_p}{\hbar\omega_p}.
    \end{split}
\end{equation*}
Successful heralds rate $J_\mathrm{herald}$ is given by the product of intraring photon pair generation rate $R$ with the fraction of pairs which exit the cavity,\\
\begin{equation}
    \begin{split}
    J_\mathrm{herald} &= R \left(\frac{\kappa_{a,\mathrm{ex}}}{\kappa_{a}}\right)^2 \\
    &= 16g^2 \frac{P_p}{\hbar\omega_p} \frac{\eta_b(1-\eta_b)}{\kappa_{b,0}} \frac{\eta^2_a(1-\eta_a)}{\kappa_{a,0}}.
    \end{split}
    \label{J_herald}
\end{equation}

\subsection{Purity Calculation}
The interaction Hamiltonian of
the $\chi(2)$-process for a low pump field, results in the an approximate biphoton of quantum state,
\begin{equation*}
    \ket{\Psi_\mathrm{II}} = \int \int d\omega_s d\omega_i \phi(\omega_s, \omega_i) \left[ a_S^\dag (\omega_s)\ket{0} \otimes a_I^\dag(\omega_i) \ket{0}\right]
\end{equation*}
Where $\phi(\omega_s, \omega_i)$ is the Joint Spectral Amplitude (JSA). The detection of a signal photon heralds the existence of an idler photon, the quantum state of which is in general an incoherent mixture of single photon amplitudes within the idler resonance. The density matrix $\rho_I$ of the heralded idler photon, defined as the partial trace of the total biphoton density matrix $\rho_{SI} = \ket{\Psi_\mathrm{II}}\bra{\Psi_\mathrm{II}}$.

\begin{equation*}
\begin{aligned}
    \rho_{SI} &= \int \int d\omega_s d\omega_i \int \int d\omega_s^{'} d\omega_i^{'} \\ 
    &\quad \phi(\omega_s, \omega_i) \phi(\omega_s^{'}, \omega_i^{'}) \left[ a_S^\dag (\omega_s)\ket{0} \otimes a_I^\dag(\omega_i) \ket{0}\right] \left[ \bra{0}a_S(\omega_s^{'}) \otimes \bra{0}a_I(\omega_i^{'})\right]\\
     \rho_I &= \mathrm{Tr}_S \left[\rho_{SI}\right]
\end{aligned}
\end{equation*}

$a_S^\dag (\omega_s)\ket{0}$ being the basis of signal frequency, partial trace of $\rho_{SI}$ is defined as

\begin{equation*}
\begin{aligned}
    \mathrm{Tr}_S \left[\rho_{SI}\right] &= \int d\omega_s^{''} \left[\bra{0}a_S (\omega_s^{''}) \otimes \mathit{I}_I\right] \rho_{SI} \left[a_S^\dag (\omega_s^{''})\ket{0} \otimes \mathit{I}_I\right]\\
    &= \int \int d\omega_s d\omega_i \int \int d\omega_s^{'} d\omega_i^{'} \phi(\omega_s, \omega_i) \phi(\omega_s^{'}, \omega_i^{'}) \\
    &\int \omega_s^{''}  \left[\bra{0}a_S (\omega_s^{''}) a_S^\dag (\omega_s)\ket{0} \otimes a_I(\omega_i)\ket{0}\right]  \left[ \bra{0}a_S (\omega_s^{'})a_S^\dag(\omega_s^{''})\ket{0} \otimes \bra{0}a_S (\omega_i^{'})\right]
\end{aligned}
\end{equation*}
For the last integral to survive, $\omega_s^{''} = \omega_s^{'} = \omega_s$ must satisfy, thus simplifying the total integral to, 

\begin{equation*}
\begin{aligned}
    \mathrm{Tr}_S \left[\rho_{SI}\right] 
    &= \int \int d\omega_i d\omega_i^{'}  \int d\omega_s \phi(\omega_s, \omega_i) \phi(\omega_s, \omega_i^{'}) \\
    & \left[\mathit{I}_S \otimes  a_I^\dag(\omega_i)\ket{0} \right]  \left[ \mathit{I}_S \otimes \bra{0}a_S(\omega_s^{'})\right] \\
    &=\int \int d\omega_i d\omega_i^{'}\phi_I(\omega_i,\omega_i^{'}) \\
    &\left[\mathit{I}_S \otimes  a_I^\dag(\omega_i)\ket{0} \right]  \left[ \mathit{I}_S \otimes \bra{0}a_S(\omega_s^{'})\right]
\end{aligned}
\end{equation*}

Defining $\phi_I(\omega_i,\omega_i^{'}) = \int d\omega_s \phi(\omega_s, \omega_i) \phi(\omega_s, \omega_i^{'})$,  purity $\gamma$ of $\rho_I$,

\begin{equation*}
\begin{aligned}
    \gamma &= \mathrm{Tr}_I \left[\rho_{I}^2\right]\\
    &= \int \int d\omega_i d\omega_i^{'} |\phi_I(\omega_i,\omega_i^{'})|^2
\end{aligned}
\end{equation*}

For a microring resonator-based (point-coupled to a channel waveguide??) SDPC, $\chi_2$ material based, photon pair source, the Joint Spectral Amplitude (JSA) can be expressed as \citeS{Smosley2008heralded,Sjeronimo2010theory},
\begin{equation*}
\begin{aligned}
    \phi(\omega_s,\omega_i) \propto \alpha_P(\omega_s+\omega_i)\mathrm{l}_P(\omega_s+\omega_i) \mathrm{l}_S(\omega_s) \mathrm{l}_I(\omega_i) 
\end{aligned}
\end{equation*}
where $\alpha_P(\omega_s+\omega_i)$ and $\mathrm{l}_x(\omega_x)$ are the pump spectral amplitude and line shape functions for corresponding resonance.\\ \\

For single pulse scheme, pulse envelope is
\begin{equation}
\begin{split}
    \alpha_\mathrm{SP}(t) &= \exp{\left[-2\ln(2) \left(\frac{t}{\tau_p}\right)^2\right]}\\
    \alpha_\mathrm{SP}(\omega) &\propto \exp{\left[-\frac{\omega^2\tau_p^2}{8\ln(2) }\right]}
\end{split}
\end{equation}

For dual pulse scheme \citeS{Schristensen2018engineering,Sburridge2020high}, pulse envelope is
\begin{equation}
\begin{split}
    \alpha_\mathrm{DP}(t) &= \sqrt{r_p}\exp{\left[-2\ln(2) \left(\frac{t}{\tau_p}\right)^2\right]} - \sqrt{1-r_p}\exp{\left[-2\ln(2) \left(\frac{(t-\Delta\tau}{\tau_p}\right)^2\right]}\\
    \alpha_\mathrm{DP}(\omega) &\propto (\sqrt{r_p} - e^{-i\Delta\tau\omega}\sqrt{1-r_p})\exp{\left[-\frac{\omega^2\tau_p^2}{8\ln(2) }\right]}
\end{split}
\end{equation}

Here, $\tau_p$ is the intensity FWHM width of the pulse, $\Delta\tau$ is the  inter-pulse temporal separation, and $r_p$ is the relative pulse weight.

\section{Squeezing Light Generation in $\chi^{2}$ microring resonator}
The Hamiltonian for squeezed light generation of a single mode in the pump-depletion assumption and in the pump reference frame is given by: 

\begin{equation}
    \mathcal{H} = \delta_a a^\dagger a + g |\beta|\left[\left(a^\dagger\right)^2+\left(a\right)^2\right] 
\end{equation}

The Linbald equation of motion for the operator $a$ with input noise operation is given by:
\begin{equation} \label{dadt}
\begin{split}
    \frac{d a}{d t} & = i[H, a] - \frac{\kappa}{2} a + \sqrt{\kappa} a_\mathrm{in} \\
    & = i\delta_a \left[a^\dagger a, a\right]+ig|\beta|^2\left[\left(a^\dagger\right)^2, a\right]-\frac{\kappa}{2} a +\sqrt{\kappa} a_\mathrm{in} \\ 
    & = -i\delta_a a - 2ig|\beta| a^\dagger - \frac{\kappa}{2} a +\sqrt{\kappa} a _\mathrm{in} \\
    & = \left[-i\delta_a -\frac{\kappa}{2}\right]a-2ig|\beta|a^\dagger+\sqrt{\kappa}a_\mathrm{in}
\end{split}
\end{equation}

Similarly, the equation of motion for $a^\dagger$ is given by:

\begin{equation} \label{daddt}
    \frac{d a^\dagger}{d t} = i\delta_a a^\dagger + 2ig|\beta|a-\frac{\kappa}{2} a^\dagger + \sqrt{\kappa} a^\dagger_\mathrm{in}
\end{equation}
Now taking Fourier Transform of eqn (\ref{dadt}),
\begin{equation} \label{FTdadt}
    \begin{split}
        & -i\omega a(\omega)  = -i\delta_a a(\omega) - 2ig|\beta|a^\dagger(-\omega) - \frac{\kappa}{2} a(\omega) +\sqrt{\kappa}a_\mathrm{in}(\omega) \\
        \Rightarrow & \left[i\left(\delta_a - \omega\right)+\frac{\kappa}{2}\right] a(\omega) +2ig|\beta|a^\dagger(-\omega) = \sqrt{\kappa}a_\mathrm{in}(\omega)
    \end{split}
\end{equation}
Similarly, from eqn (\ref{daddt}),
\begin{equation}\label{FTdaddt}
    \left[-i\left(\delta_a + \omega\right)+\frac{\kappa}{2}\right] a^\dagger(-\omega) - 2ig|\beta|a(\omega) = \sqrt{\kappa}a^\dagger_\mathrm{in}(-\omega)
\end{equation}
In matrix form, the equations (\ref{FTdadt}) and (\ref{FTdaddt}) become:

\begin{equation}
    \begin{pmatrix}
        \frac{\kappa}{2}+i(\delta_a - \omega) & i2g|\beta| \\
        -i2g|\beta| & \frac{\kappa}{2}- i(\delta_a+\omega)
    \end{pmatrix} \begin{pmatrix}
        a(\omega) \\ a^\dagger(-\omega)
    \end{pmatrix} = \sqrt{\kappa}\begin{pmatrix}
        a_\mathrm{in}(\omega) \\ a^\dagger_\mathrm{in}(-\omega) 
    \end{pmatrix}
\end{equation}
Solving this equation,
\begin{equation}
    a(\omega) = \frac{\sqrt{\kappa}\left(\frac{\kappa}{2}-i\left(\delta_a+\omega\right)\right)a_\mathrm{in}(\omega)-i2\sqrt{\kappa}g|\beta|a^\dagger_\mathrm{in}(-\omega)}{\Delta}
\end{equation}
Here, $\Delta = \left[\left(\frac{\kappa}{2}+i\left(\delta_a-\omega\right)\right)\left(\frac{\kappa}{2}-i\left(\delta_a+\omega\right)\right)\right]-4g^2|\beta|^2$. Now the output mode is linearly related to the input vacuum mode as:
\begin{equation*}
    a_\mathrm{out}(\omega) = \sqrt{\kappa}a(\omega) - a_\mathrm{in}(\omega)
\end{equation*}
So, the output mode is given by:
\begin{equation}
    \begin{split}
        a_\mathrm{out}(\omega) & = \frac{\kappa\left(\frac{\kappa}{2}-i\left(\delta_a+\omega\right)\right)a_\mathrm{in}-i2g\kappa|\beta|a^\dagger_\mathrm{in}(-\omega)}{\Delta} - a_\mathrm{in}(\omega) \\
        & = \frac{\left[\kappa\left(\frac{\kappa}{2}-i\left(\delta_a+\omega\right)\right)-\left(\frac{\kappa}{2}+i\left(\delta_a-\omega\right)\right)\left(\frac{\kappa}{2}-i\left(\delta_a+\omega\right)\right)+4g^2|\beta|^2\right]a_\mathrm{in}(\omega)-i2g\kappa|\beta|a^\dagger_\mathrm{in}(-\omega)}{\Delta} \\
        & = \frac{\left[\left(\frac{\kappa}{2}-i\left(\delta_a+\omega\right)\right)\left(\frac{\kappa}{2}-i\left(\delta_a-\omega\right)\right)+4g^2|\beta|^2\right]a_\mathrm{in}(\omega)-i2g\kappa|\beta|a^\dagger_\mathrm{in}(-\omega)}{\Delta}
    \end{split}
\end{equation}
Now if the detuning $\delta_a = 0$,
\begin{equation*}
    a_\mathrm{out}(\omega) = \frac{\left(\frac{\kappa^2}{4}+\omega^2+4g^2|\beta|^2\right)a_\mathrm{in}(\omega)-i2g\kappa|\beta|a^\dagger_\mathrm{in}(-\omega)}{\left(\frac{\kappa}{2}-i\omega\right)^2-4g^2|\beta|^2}
\end{equation*}
Finally, in the presence of non-zero detuning,
\begin{equation}
\begin{split}
    a_\mathrm{out}(\omega) & = \frac{\left(\frac{\kappa}{2}-i\left(\delta_a+\omega\right)\right)\left(\frac{\kappa}{2}-i\left(\delta_a-\omega\right)\right)+4g^2|\beta|^2}{\Delta}a_\mathrm{in}(\omega)-i\frac{2g\kappa|\beta|}{\Delta}a^\dagger_\mathrm{in}(-\omega) \\
    & = \frac{\left(\frac{\kappa}{2}-i\delta_a\right)^2+\omega^2+4g^2|\beta|^2}{\left(\frac{\kappa}{2}-i\omega\right)^2+\delta^2_a-4g^2|\beta|^2}a_\mathrm{in}(\omega) - i\frac{2g\kappa|\beta|}{\left(\frac{\kappa}{2}-i\omega\right)^2+\delta^2_a-4g^2|\beta|^2}a^\dagger_\mathrm{in}(-\omega) \\
a^\dagger_\mathrm{out} (\omega) & = \frac{\left(\frac{\kappa}{2}+i\delta_a\right)^2+\omega^2+4g^2|\beta|^2}{\left(\frac{\kappa}{2}+i\omega\right)^2+\delta^2_a-4g^2|\beta|^2} a^\dagger_\mathrm{in}(\omega) + i \frac{2g\kappa|\beta|}{\left(\frac{\kappa}{2}+i\omega\right)^2+\delta^2_a-4g^2|\beta|^2} a_\mathrm{in}(-\omega)    
\end{split}
\end{equation}

The quadrature operators of the output modes are defined as: $X_\mathrm{out, 1} = a_\mathrm{out}+a^\dagger_\mathrm{out}$ and $X_\mathrm{out, 2} = i\left(a_\mathrm{out}-a^\dagger_\mathrm{out}\right)$. For squeezing level calculation, we need to know the variance of the output modes as the squeezed modes are defined in terms of their variances. The variance of $X_\mathrm{1, out}$ is $\langle\Delta X_\mathrm{1, out}\rangle = \langle X_\mathrm{1, out}(\omega)X_\mathrm{1, out}(\omega^\prime)\rangle - \langle X_\mathrm{1, out}(\omega) \rangle\langle X_\mathrm{1, out}(\omega^\prime) \rangle$. However, $\langle X_\mathrm{1, out} \rangle \sim \langle a_\mathrm{out} + a^\dagger_\mathrm{out} \rangle \sim 0$ as $\langle a_\mathrm{in}(\omega) \rangle = 0$ because this is vacuum noise input. We also define, $\hat{a}_\mathrm{out}(\omega) =  e^{i\phi_\mathrm{out}}\hat{a}_\mathrm{out}(\omega)$. Then $\Delta X_\mathrm{1, out}$ can be written as
\begin{equation}
\begin{split}
     \langle \Delta X_\mathrm{1, out} \rangle & = \langle X_\mathrm{1, out}(\omega) X_\mathrm{1, out}(\omega^\prime) \\
     & = \Big\langle \left(a_\mathrm{out}(\omega)+a^\dagger_\mathrm{out}(\omega)\right)\left(a_\mathrm{out}(\omega^\prime)+a^\dagger_\mathrm{out}(\omega^\prime)\right)\Big\rangle \\
     & = \langle a_\mathrm{out}(\omega)a_\mathrm{out}(\omega^\prime)+a_\mathrm{out}^\dagger(\omega)a_\mathrm{out}(\omega^\prime)+ \\
     & a_\mathrm{out}(\omega)a_\mathrm{out}^\dagger(\omega^\prime)+a_\mathrm{out}(\omega^\prime)a_\mathrm{out}(\omega)\rangle \\
     & = e^{2i\phi_\mathrm{out}}\langle a_\mathrm{out}(\omega)a_\mathrm{out}(\omega^\prime)\rangle + e^{-2i\phi_\mathrm{out}}\langle a_\mathrm{out}^\dagger(\omega)a_\mathrm{out}^\dagger(\omega^\prime)\rangle + \\
     & 2\langle  a_\mathrm{out}^\dagger(\omega)a_\mathrm{out}(\omega^\prime)\rangle + \left[a_\mathrm{out}(\omega), a_\mathrm{out}^\dagger(\omega^\prime)\right] \\
     & = e^{2i\phi_\mathrm{out}}\langle a_\mathrm{out}(\omega)a_\mathrm{out}(\omega^\prime)\rangle + e^{-2i\phi_\mathrm{out}}\langle a_\mathrm{out}^\dagger(\omega)a_\mathrm{out}^\dagger(\omega^\prime)\rangle + \\
     & 2\langle  a^\dagger_\mathrm{out}(\omega)a_\mathrm{out}(\omega^\prime)\rangle + \delta\left(\omega-\omega^\prime\right) 
\end{split}
\end{equation}
Here, $\left[a_\mathrm{out}(\omega), a_\mathrm{out}^\dagger(\omega^\prime)\right] = \delta\left(\omega-\omega^\prime\right)$ Now, 
\begin{equation} \label{eqaa}
    \begin{split}
        \langle a_\mathrm{out}(\omega)a_\mathrm{out}(\omega^\prime)\rangle & = -i \frac{2g\kappa|\beta|\left[\left(\frac{\kappa}{2}-i\delta_a\right)^2+\omega^2+4g^2|\beta|^2\right]}{\left[\left(\frac{\kappa}{2}-i\omega\right)^2+\delta^2_a-4g^2|\beta|^2\right]\left[\left(\frac{\kappa}{2}-i\omega^\prime\right)^2+\delta^2_a-4g^2|\beta|^2\right]}\delta(\omega+\omega^\prime)
    \end{split}
\end{equation}

\begin{equation}\label{eqadad}
    \begin{split}
        \langle a^\dagger_\mathrm{out}(\omega)a^\dagger_\mathrm{out}(\omega^\prime)\rangle & = i \frac{2g\kappa|\beta|\left[\left(\frac{\kappa}{2}+i\delta_a\right)^2+\omega^{\prime2}+4g^2|\beta|^2\right]}{\left[\left(\frac{\kappa}{2}+i\omega\right)^2+\delta^2_a-4g^2|\beta|^2\right]\left[\left(\frac{\kappa}{2}+i\omega^\prime\right)^2+\delta^2_a-4g^2|\beta|^2\right]}\delta(\omega+\omega^\prime)
    \end{split}
\end{equation}

\begin{equation}\label{eqada}
    \begin{split}
        \langle a^\dagger_\mathrm{out}(\omega)a_\mathrm{out}(\omega^\prime)\rangle & =  \frac{4g^2\kappa^2|\beta|^2}{\left[\left(\frac{\kappa}{2}+i\omega\right)^2+\delta^2_a-4g^2|\beta|^2\right]\left[\left(\frac{\kappa}{2}-i\omega^\prime\right)^2+\delta^2_a-4g^2|\beta|^2\right]}\delta(\omega-\omega^\prime)
    \end{split}
\end{equation}

Now using eqn (\ref{eqaa}), (\ref{eqadad}) and (\ref{eqada}) and integrating the equation over all possible $\omega^\prime$,

\begin{equation}
\begin{split}
    S_\mathrm{XX} (\omega) & = e^{2i\phi_\mathrm{out}} \int_{-\infty}^{\infty} \langle a_\mathrm{out}\left(\omega\right)a_\mathrm{out}\left(\omega^\prime\right)\rangle d\omega^\prime + e^{-2i\phi_\mathrm{out}} \int_{-\infty}^{\infty} \langle a^\dagger_\mathrm{out}\left(\omega\right)a^\dagger_\mathrm{out}\left(\omega^\prime\right)\rangle d\omega^\prime + \\
    & 2 \int_{-\infty}^{\infty} \langle a^\dagger_\mathrm{out}\left(\omega\right)a_\mathrm{out}\left(\omega^\prime\right)\rangle d\omega^\prime + \int_{-\infty}^{\infty} \delta\left(\omega-\omega^\prime\right)d\omega^\prime \\
    & = 1-e^{2i\phi_\mathrm{out}}i \frac{2g\kappa|\beta|\left[\left(\frac{\kappa}{2}-i\delta_a\right)^2+\omega^2+4g^2|\beta|^2\right]}{\left[\left(\frac{\kappa}{2}-i\omega\right)^2+\delta^2_a-4g^2|\beta|^2\right]\left[\left(\frac{\kappa}{2}+i\omega\right)^2+\delta^2_a-4g^2|\beta|^2\right]}+ \\
    & e^{-2i\phi_\mathrm{out}}i \frac{2g\kappa|\beta|\left[\left(\frac{\kappa}{2}+i\delta_a\right)^2+\omega^{2}+4g^2|\beta|^2\right]}{\left[\left(\frac{\kappa}{2}+i\omega\right)^2+\delta^2_a-4g^2|\beta|^2\right]\left[\left(\frac{\kappa}{2}-i\omega\right)^2+\delta^2_a-4g^2|\beta|^2\right]}+ \\
    & 2\frac{4g^2\kappa^2|\beta|^2}{\left[\left(\frac{\kappa}{2}+i\omega\right)^2+\delta^2_a-4g^2|\beta|^2\right]\left[\left(\frac{\kappa}{2}-i\omega\right)^2+\delta^2_a-4g^2|\beta|^2\right]}
\end{split}
\end{equation}

For the simplicity of calculation, let us assume, $\delta_a = 0$.
\begin{equation}
\begin{split}
    S_\mathrm{XX}(\omega) & = 1-e^{2i\phi_\mathrm{out}}i \frac{2g\kappa|\beta|\left[\frac{\kappa^2}{4}+\omega^2+4g^2|\beta|^2\right]}{\left[\left(\frac{\kappa}{2}-i\omega\right)^2-4g^2|\beta|^2\right]\left[\left(\frac{\kappa}{2}+i\omega\right)^2-4g^2|\beta|^2\right]}+ \\
    & e^{-2i\phi_\mathrm{out}}i \frac{2g\kappa|\beta|\left[\frac{\kappa^2}{4}+\omega^{2}+4g^2|\beta|^2\right]}{\left[\left(\frac{\kappa}{2}+i\omega\right)^2-4g^2|\beta|^2\right]\left[\left(\frac{\kappa}{2}-i\omega\right)^2-4g^2|\beta|^2\right]}+ \\
    & 2\frac{4g^2\kappa^2|\beta|^2}{\left[\left(\frac{\kappa}{2}+i\omega\right)^2-4g^2|\beta|^2\right]\left[\left(\frac{\kappa}{2}-i\omega\right)^2-4g^2|\beta|^2\right]} \\
    & = 1 + \frac{2g\kappa|\beta|\left[\frac{\kappa^2}{4}+\omega^2+4g^2|\beta|^2\right]}{\left[\left(\frac{\kappa}{2}-i\omega\right)^2-4g^2|\beta|^2\right]\left[\left(\frac{\kappa}{2}+i\omega\right)^2-4g^2|\beta|^2\right]}\left(ie^{-2i\phi_\mathrm{out}}-ie^{2i\phi_\mathrm{out}}\right) + \\
    & 2\frac{4g^2\kappa^2|\beta|^2}{\left[\left(\frac{\kappa}{2}+i\omega\right)^2-4g^2|\beta|^2\right]\left[\left(\frac{\kappa}{2}-i\omega\right)^2-4g^2|\beta|^2\right]} \\
    & = 1+2\sin{\phi_\mathrm{out}}\frac{2g\kappa|\beta|\left[\frac{\kappa^2}{4}+\omega^2+4g^2|\beta|^2\right]}{\left[\left(\frac{\kappa}{2}-i\omega\right)^2-4g^2|\beta|^2\right]\left[\left(\frac{\kappa}{2}+i\omega\right)^2-4g^2|\beta|^2\right]} + \\
    & 2\frac{4g^2\kappa^2|\beta|^2}{\left[\left(\frac{\kappa}{2}+i\omega\right)^2-4g^2|\beta|^2\right]\left[\left(\frac{\kappa}{2}-i\omega\right)^2-4g^2|\beta|^2\right]} 
\end{split}
\end{equation}

Now $S_\mathrm{XX}(\omega)$ is maximum when $\sin{\phi_\mathrm{out}} = 1$ and minimum when $\sin\left({\phi_\mathrm{out}}\right) = -1$. We denote maximum (minimum) value of $S_\mathrm{XX}(\omega)$ as $S_{+}(\omega)$($S_{-}(\omega)$).
Therefore,
\begin{equation}
\begin{split}
    S_+(\omega) & = 1+\frac{4g\kappa|\beta|\left[\frac{\kappa^2}{4}+\omega^2+4g^2|\beta|^2\right]}{\left[\left(\frac{\kappa}{2}-i\omega\right)^2-4g^2|\beta|^2\right]\left[\left(\frac{\kappa}{2}+i\omega\right)^2-4g^2|\beta|^2\right]} + \\
    & 2\frac{4g^2\kappa^2|\beta|^2}{\left[\left(\frac{\kappa}{2}+i\omega\right)^2-4g^2|\beta|^2\right]\left[\left(\frac{\kappa}{2}-i\omega\right)^2-4g^2|\beta|^2\right]} \\ 
    & = 1+\frac{4g\kappa|\beta|\left[\frac{\kappa^2}{4}+\omega^2+4g^2|\beta|^2\right]}{\left(\left(\frac{\kappa}{2}+2g|\beta|\right)^2+\omega^2\right)\left(\left(\frac{\kappa}{2}-2g|\beta|\right)^2+\omega^2\right)}+ \\
    &\frac{8g^2\kappa^2|\beta|^2}{\left(\left(\frac{\kappa}{2}+2g|\beta|\right)^2+\omega^2\right)\left(\left(\frac{\kappa}{2}-2g|\beta|\right)^2+\omega^2\right)} \\
    & = 1+4g\kappa|\beta|\frac{\frac{\kappa^2}{4}+\omega^2+4g^2|\beta|^2+2g\kappa|\beta|}{\left(\left(\frac{\kappa}{2}+2g|\beta|\right)^2+\omega^2\right)\left(\left(\frac{\kappa}{2}-2g|\beta|\right)^2+\omega^2\right)} \\
    & = 1+\frac{4g\kappa|\beta|}{\left(\frac{\kappa}{2}-2g|\beta|\right)^2+\omega^2}
\end{split}    
\end{equation}
Similarly,
\begin{equation}
    S_-(\omega) = 1-\frac{4g\kappa|\beta|}{\left(\frac{\kappa}{2}+2g|\beta|\right)^2+\omega^2}
\end{equation}
So, $S_\pm$ ($\omega$) can be written compactly as \citeS{Spark2024single,Sarnbak2019compact}:
\begin{equation}
        S_\pm(\omega) = 1\pm\frac{4g\kappa|\beta|}{\left(\frac{\kappa}{2}\mp2g|\beta|\right)^2+\omega^2} = 1\pm\frac{4g\kappa|\beta|}{\left(\frac{\kappa}{2}\mp2g|\beta|\right)^2+\omega^2}
\end{equation}

The squeezed (anti-squeezed) state can be modulated by vacuum shot noise. If $p_1$ is the probability of the field quadrature spectrum $S_\mathrm{XX, out}$ being detected and $p_2$ is the probability of shot noise being detected, the resultant spectrum density can be written as:
\begin{equation}
    \begin{split}
     S_\mathrm{XX,detected} & = p_1S_\mathrm{XX, out} + p_2S_\mathrm{XX, vac} = p_1S_\mathrm{XX, out} + \left(1-p_1\right)S_\mathrm{XX, out}\\
     & = p_1S_\mathrm{XX, out} + \left(1-p_1\right)   
    \end{split}
\end{equation}
Here, $p_1+p_2 = 1$ and $S_\mathrm{XX,vac} = 1$. 
Now, incorporating different sources of loss in the extraction and detection systems, $p_1 = \eta_\mathrm{a} \eta_\mathrm{T} \eta_\mathrm{d}$ where $\eta_\mathrm{a}$ is the coupling ratio of the resonator, $\eta_\mathrm{T}$ is the transmission from the cavity to the detector and $\eta_\mathrm{d}$ is the quantum efficiency of the detector. Finally, including these losses, the squeezing and anti-squeezing level becomes,
\begin{equation}
    S_{\pm}(\omega) = 1\pm \eta_\mathrm{a} \eta_\mathrm{T} \eta_\mathrm{d} \frac{4g\kappa |\beta|}{\left(\frac{\kappa}{2}\mp2g|\beta|\right)^2+\omega^2}
\end{equation}
In our analysis, we assume perfect transmission from the chip and detector efficiency for squeezing measurement i.e., $\eta_\mathrm{T} = 1$ and $\eta_\mathrm{d} = 1$. By defining $|\beta_\mathrm{th}| = \frac{\kappa}{4g}$ and corresponding power as $P_\mathrm{th}$, $S_{\pm}$ can be written as follows:
\begin{equation}
    \begin{split}
        S_{\pm} & = 1\pm \eta_a \frac{4 \frac{|\beta|}{|\beta_\mathrm{th}|}}{\left(1\mp\frac{|\beta|}{|\beta_\mathrm{th}|}\right)^2+4\left(\frac{\omega}{\kappa}\right)^2} \\
        & = 1\pm \eta_a \frac{4 \sqrt{ \frac{P_s}{P_\mathrm{th}}}}{\left(1\mp\sqrt{\frac{P_s}{P_\mathrm{th}}}\right)^2+4\left(\frac{\omega}{\kappa}\right)^2} 
    \end{split}
\end{equation}
Here, $\omega$ is the side-band frequency of the local oscillator. To differentiate it from angular frequency, we replace $\omega$ in the expression of $S_\pm$ by $\Delta \omega_\mathrm{LO}$. So the final expression for $S_\pm$ becomes
\begin{equation}
\begin{split}
    S_{\pm} &= 1\pm \eta_a \frac{4 \sqrt{ \frac{P_s}{P_\mathrm{th}}}}{\left(1\mp\sqrt{\frac{P_s}{P_\mathrm{th}}}\right)^2+4\left(\frac{\Delta\omega_\mathrm{LO}}{\kappa}\right)^2} \\
    &= 1 \pm \eta_a \frac{4 \sqrt{\frac{P_s}{P_\mathrm{th}}}}{ (1\mp  \sqrt{\frac{P_s}{P_\mathrm{th}}})^2 + 4(1-\eta_a)^2(\frac{\Delta\omega_\mathrm{LO}}{\kappa_{a,0}})^2}
\end{split}
\end{equation}
The pump power threshold for OPO is given by:
\begin{equation}
    \begin{split}
    P_\mathrm{th} &= \left(1 + \frac{\delta^2}{\kappa_a^2}\right)\frac{\hbar \omega_b}{64g^2} \frac{\kappa_a^2 \kappa_b^2}{\kappa_{b,ex}}\\
    &= \left(1 + \frac{\delta^2 (1-\eta_a)^2}{\kappa_{a,0}^2}\right)\frac{\hbar\omega_b}{64g^2} \frac{\kappa_{a,0}^2}{(1-\eta_a)^2} \frac{\kappa_{b,0}}{\eta_b(1-\eta_b)} 
    \end{split} 
\end{equation}
Here, $\eta$ are coupling ratio defined as $\eta_a = \frac{\kappa_{a,1}}{\kappa_{a}}$, $\eta_b = \frac{\kappa_{b,1}}{\kappa_{b}}$.

\begin{figure}[H]
    \centering
    \includegraphics[width = 0.6\textwidth]{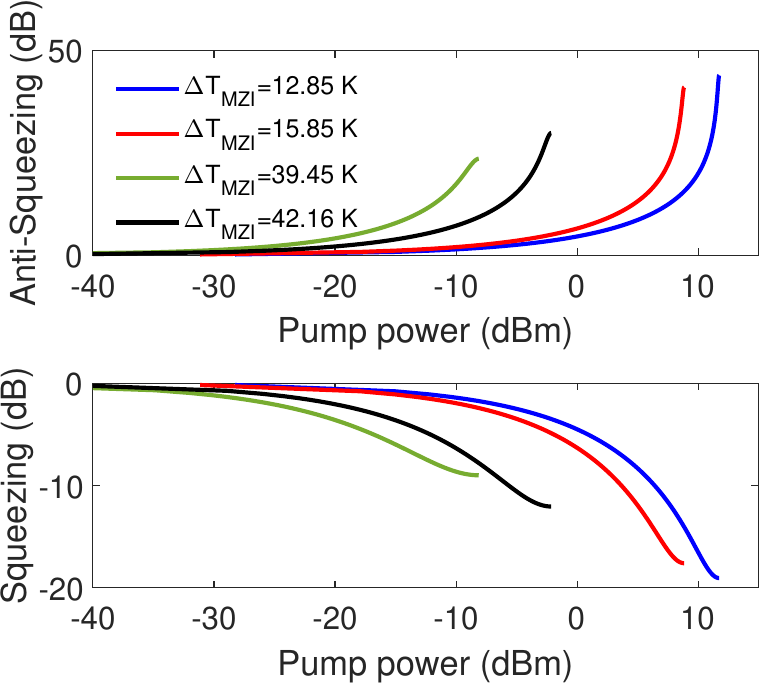}
    \caption{a) Anti-squeezing and b) squeezing levels at four different coupling ratios of the telecom mode achievable with the MZI thermal tuner with a critically coupled pump. The plots are truncated after the threshold power for each coupling ratio.}
    \label{fig:RMZI}
\end{figure}


\bibliographystyleS{unsrt}
\bibliographyS{refssuppl}

\end{document}